\documentclass[letter]{aa} 

\usepackage{graphicx}
\usepackage{txfonts}
\usepackage{gensymb}
\usepackage{xspace}
\usepackage{tabularx}
\usepackage{placeins}

\usepackage{natbib,twoopt}
\usepackage[breaklinks=true]{hyperref} 
\bibpunct{(}{)}{;}{a}{}{,} 
\makeatletter
\newcommandtwoopt{\citeads}[3][][]{\href{http://adsabs.harvard.edu/abs/#3}%
{\def\hyper@linkstart##1##2{}%
\let\hyper@linkend\@empty\citealp[#1][#2]{#3}}}
\newcommandtwoopt{\citepads}[3][][]{\href{http://adsabs.harvard.edu/abs/#3}%
{\def\hyper@linkstart##1##2{}%
\let\hyper@linkend\@empty\citep[#1][#2]{#3}}}
\newcommandtwoopt{\citetads}[3][][]{\href{http://adsabs.harvard.edu/abs/#3}%
{\def\hyper@linkstart##1##2{}%
\let\hyper@linkend\@empty\citet[#1][#2]{#3}}}
\newcommandtwoopt{\citeyearads}[3][][]%
{\href{http://adsabs.harvard.edu/abs/#3}
{\def\hyper@linkstart##1##2{}%
\let\hyper@linkend\@empty\citeyear[#1][#2]{#3}}}
\makeatother
\def\ms{\hbox{m\,s$^{-1}$}}         
\def\m2s2{\hbox{\,m$^{2}$\,s$^{-2}$}} 
\def\kms{\hbox{\,km\,s$^{-1}$}}       
\def\vsini{\hbox{$v$\,sin\,$i_{\star}$}}      
\def\Msun{$M_{\odot}$\xspace}             
\def\Rsun{$R_{\odot}$\xspace}

\def\ten[#1]{$\;\times 10^{#1}$}

\def\logg{$\log g$}

\newcommand{\Rnom}{\hbox{$\mathcal{R}^{\rm N}_{\odot}$}} 

\newcommand{\GMnom}{\hbox{$\mathcal{(GM)}^{\rm N}_{\odot}$}}

\newcommand{\Renom}{\hbox{$\mathcal{R}^{\rm N}_{e \rm E}$}}

\newcommand{\GMenom}{\hbox{$\mathcal{(GM)}^{\rm N}_{\rm E}$}}

\newcommand{\emcee}{{\sc \tt emcee}\xspace}
\newcommand{\juliet}{{\sc \tt juliet}\xspace}
\newcommand{\batman}{{\sc \tt batman}\xspace}
\newcommand{\radvel}{{\sc \tt radvel}\xspace}

\newcommand{\dynesty}{{\sc \tt dynesty}\xspace}

\newcommand{\specmatch}{{\sc \tt SpecMatch-Emp}\xspace}
\newcommand{\starry}{{\sc \tt starry}\xspace}
\newcommand{\pymc}{{\sc \tt pymc3}\xspace}

\newcommand{\REarth}{$\mathrm{R_E}$\xspace}
\newcommand{\MEarth}{$\mathrm{M_E}$\xspace}

\begin{document} 

   \title{TOI-3884~b: A rare 6-\REarth planet that transits a low-mass star with a giant and likely polar spot} 

   \author{J.M.~Almenara\inst{\ref{grenoble}}
        \and X.~Bonfils\inst{\ref{grenoble}}
        \and T.~Forveille\inst{\ref{grenoble}}
        \and N.~Astudillo-Defru\inst{\ref{concepcion}}
        \and D.R.~Ciardi\inst{\ref{caltech}} 
        \and R.P.~Schwarz\inst{\ref{harvard}} 
        \and K.A.~Collins\inst{\ref{harvard}} 
        \and M.~Cointepas\inst{\ref{grenoble},\ref{geneva}}
        \and M.B.~Lund\inst{\ref{caltech}} 
        \and F.~Bouchy\inst{\ref{geneva}}
        \and D.~Charbonneau\inst{\ref{harvard}} 
        \and R.F.~D\'{i}az\inst{\ref{ba}}
        \and X.~Delfosse\inst{\ref{grenoble}}
        \and R.C.~Kidwell\inst{\ref{stsi}} 
        \and M.~Kunimoto\inst{\ref{mit}} 
        \and D.W.~Latham\inst{\ref{harvard}} 
        \and J.J.~Lissauer\inst{\ref{ames}}
        \and F.~Murgas\inst{\ref{iac},\ref{ull}}
        \and G.~Ricker\inst{\ref{mit}} 
        \and S.~Seager\inst{\ref{mit},\ref{mit2},\ref{mit3}} 
        \and M.~Vezie\inst{\ref{mit}} 
        \and D.~Watanabe\inst{\ref{Fredericksburg}} 
      }

      \institute{
        Univ. Grenoble Alpes, CNRS, IPAG, F-38000 Grenoble, France\label{grenoble}
        \and Departamento de Matem\'{a}tica y F\'{i}sica Aplicadas, Universidad Cat\'{o}lica de la Sant\'{i}sima Concepci\'{o}n, Alonso de Rivera 2850, Concepci\'{o}n, Chile\label{concepcion}
        \and NASA Exoplanet Science Institute, Caltech/IPAC, Pasadena, CA 91125, USA\label{caltech}
        \and Center for Astrophysics \textbar \ Harvard \& Smithsonian, 60 Garden Street, Cambridge, MA 02138, USA\label{harvard}  
        \and Observatoire de Gen\`eve, Département d’Astronomie, Universit\'e de Gen\`eve, Chemin Pegasi 51, 1290 Versoix, Switzerland\label{geneva}
        \and International Center for Advanced Studies (ICAS) and ICIFI (CONICET), ECyT-UNSAM, Campus Miguelete, 25 de Mayo y Francia, (1650) Buenos Aires, Argentina\label{ba}
        \and Space Telescope Science Institute, 3700 San Martin Drive, Baltimore, MD, 21218, USA\label{stsi}        
        \and Department of Physics and Kavli Institute for Astrophysics and Space Research, Massachusetts Institute of Technology, Cambridge, MA 02139, USA\label{mit}
        \and NASA Ames Research Center, Moffett Field, CA 94035, USA\label{ames}
        \and Instituto de Astrofísica de Canarias (IAC), E-38200 La Laguna, Tenerife, Spain\label{iac}
        \and Dept. Astrofísica, Universidad de La Laguna (ULL), E-38206 La Laguna, Tenerife, Spain\label{ull}
        \and Department of Earth, Atmospheric and Planetary Sciences, Massachusetts Institute of Technology, Cambridge, MA 02139, USA\label{mit2}
        \and Department of Aeronautics and Astronautics, MIT, 77 Massachusetts Avenue, Cambridge, MA 02139, USA\label{mit3}
        \and Planetary Discoveries in Fredericksburg, VA 22405, USA\label{Fredericksburg}
        }
      \date{}

      \date{}

 
  \abstract
   {
The Transiting Exoplanet Survey Satellite mission identified a deep and asymmetric transit-like signal with a periodicity of 4.5~days orbiting the M4 dwarf star TOI-3884. The signal has been confirmed by follow-up observations collected by the ExTrA facility and Las Cumbres Observatory Global Telescope, which reveal that the transit is chromatic. The light curves are well modelled by a host star having a large polar spot transited by a 6-\REarth planet. We validate the planet with seeing-limited photometry, high-resolution imaging, and radial velocities. TOI-3884~b, with a radius of $6.00 \pm 0.18$~\REarth, is the first sub-Saturn planet transiting a mid-M dwarf. Owing to the host star's brightness and small size, it has one of the largest transmission spectroscopy metrics for this planet size and becomes a top target for atmospheric characterisation with the James Webb Space Telescope and ground-based telescopes.   
   }

   \authorrunning{J.M. Almenara et al.}
   \titlerunning{TOI-3884}

   \maketitle
%

\section{Introduction}

Over the past thirty years, more than 3{,}500 transiting planets have been discovered and important statistical properties have emerged. For example, it has been found that planets larger than Neptune are uncommon around later type stars. The Bern population synthesis model, for instance, predicts that no planet larger than 5~\REarth exists around stars with $M_\star < 0.3$~\Msun \citep{burn2021}. And indeed, no planet larger than $\sim 3.5$~\REarth has been found so far around any star with $R_\star \lesssim 0.35$~\Rsun. One may note an apparent exception to this statistic: the radial-velocity (RV) detection of a giant planet around the small star GJ\,3512 (0.14~\Rsun). With a minimum mass of $M_p \sin i \sim 147$~\MEarth, GJ\,3512\,b is presumably larger than $\sim 10$~\REarth \citep{morales2019}. GJ\,3512\,b is, however, found beyond the ice line ($a = 0.34$~au) of its star, and \citet{burn2021} could, although with difficulty, form a similar planet by slowing down planet migration in their models. That model tuning would, however, not work for a giant planet closer to the star, as its migration could not be turned down.

TOI-3884.01, a planet candidate identified by the Transiting Exoplanet Survey Satellite \citep[TESS,][]{ricker2015} mission, is also at odds with this statistic, and is another potential challenge to planet formation models. TOI-3884.01 would be 6~\REarth in radius and orbiting at $0.035$~au from a 0.30~\Rsun star. This is above the largest predicted size for planets around such stars, especially at a small separation.

In addition, the transits of TOI-3884.01 are asymmetric, which we show can be explained by a giant polar spot on the surface of its fully convective host star. Polar spots do not exist on the Sun but their existence has been proposed since the first Doppler image of stars with short rotation periods \citep{vogt1983,strassmeier1996}. Concerning M dwarfs, several studies using spectropolarimetric observations \citep{donati2006,morin2008,moutou2017} have demonstrated that mid-M dwarfs with short rotation periods frequently develop a dipolar and axisymmetric magnetic field. \citet{hebrard2016} show that, for this spectral type, spots are frequently concentrated at the magnetic pole, and this suggests that polar spots may be frequent. 

In this Letter we report the confirmation of the sub-Saturn planet TOI-3884~b transiting an M4 dwarf. This planet is the best target of its size for atmospheric characterisation.

\section{Observations}\label{section.observations}

\subsection{TESS}\label{sec.tess}

The TESS mission observed TOI-3884 (TIC 86263325) in sector 22 with a 30-minute cadence\footnote{The photometry is available in the full-frame images (FFIs) that were calibrated by the SPOC at NASA's Ames Research Center \citep{jenkins2016}.}, and in sectors 46 and 49 with a 2-minute cadence, for a total of 81~days spanning 765~days. A transit signature with a 4.545-day period and 4\% depth was first identified by the Faint Star Search \citep{kunimoto2022} using data products from the Quick-Look Pipeline \citep[QLP,][]{huang2020,kunimoto2021}. Gaia has detected no other star within the TESS aperture \citep[Fig.~\ref{fig.tpfplotter},][]{aller2020,gaia2018}. The TESS Science Processing Operations Centre (SPOC) Simple Aperture Photometry \citep[SAP,][]{twicken2010,morris2020} and the QLP Kepler Spline SAP \citep[KSPSAP,][]{vanderburg2014} light curves are shown in Fig.~\ref{fig.tess}. The SPOC SAP light curve shows no signs of rotational modulation. For the analysis in Sect.~\ref{section.analysis}, we used the KSPSAP (sector 22) and the Pre-search Data Conditioning (PDC) SAP \citep[PDCSAP,][sectors 46 and 49]{strumpe2012,strumpe2014,smith2012} light curves in which long-term trends have been removed. TESS observed a total of 13 transits and Fig.~\ref{fig.tess} shows one composite transit per sector. The asymmetry of the transit-like signature in sector 46 triggered our follow-up observations with the Exoplanets in Transits and their Atmospheres (ExTrA) project.

\subsection{Near-IR photometry with ExTrA}

The ExTrA facility \citep{bonfils2015}, located at La Silla observatory, consists of a near-IR (0.88 to 1.55~$\mu$m) multi-object spectrograph fibre-fed by three 60-cm telescopes. Five fibre positioners intercept the light from one target and four comparison stars at the focal plane of each telescope. We observed five transits of TOI-3884~b on nights UTC 2022 February 22, 2022 March 3, 2022 April 4, 2022 April 13, and 2022 May 15. We observed with one, two, or three telescopes simultaneously. We used the 8\arcsec\ aperture fibres, the low-resolution mode of the spectrograph (R$\sim$20), and 60-second exposures. We chose comparison stars with effective temperatures \citep{gaia2018} similar to that of TOI-3884. The ExTrA data were analysed using custom data reduction software.

\subsection{Seeing-limited optical photometry}

We obtained one additional ground-based photometric follow-up observation of TOI-3884 with the Las Cumbres Observatory Global Telescope \citep[LCOGT;][]{Brown:2013} and as part of the {\em TESS} Follow-up Observing Program\footnote{\url{https://tess.mit.edu/followup}} \citep[TFOP;][]{collins:2019}, in an attempt to rule out or identify nearby eclipsing binaries (NEBs) as potential sources of the \textit{TESS} detection, to measure the transit-like event on target to confirm its depth, and thus the \textit{TESS} photometric deblending factor, and to refine the \textit{TESS} ephemeris. We scheduled our transit observations using the {\tt TESS Transit Finder}, which is a customised version of the {\tt Tapir} software package \citep{Jensen:2013}.

The LCOGT observation was performed with a 1-m telescope at Teide observatory in the g' filter, with an exposure time of 300~seconds. The data were calibrated by the standard LCOGT {\tt BANZAI} pipeline \citep{McCully:2018}, and the photometry was extracted using {\tt AstroImageJ} \citep{Collins:2017}. The transit of TOI-3884~b was detected on-target within the photometric aperture of 4.7\arcsec.

\subsection{High-resolution imaging}
As part of our standard process for validating transiting exoplanets and assessing the possible contamination of the derived planetary radii by bound or unbound stars in the TESS aperture \citep{ciardi2015}, we obtained near-IR adaptive optics (AO) imaging of TOI-3884 at Palomar Observatory. The Palomar Observatory observations of TOI-3884 were made in the narrow-band Br-$\gamma$ filter $(\lambda_o = 2.1686; \Delta\lambda = 0.0326~\mu$m) with the PHARO instrument \citep{hayward2001} behind the P3K natural guide star AO system \citep{dekany2013} on 2022~February~13 in a standard 5-point quincunx dither pattern with steps of 5\arcsec. Each dither position was observed three times, offset in position from each other by 0.5\arcsec\ for a total of 15 frames; the individual exposures were 31.1 seconds long, for a total on-source times of 466~seconds. PHARO has a pixel scale of $0.025\arcsec$ per pixel for a total field of view of $\sim25\arcsec$.
    
The AO data were processed and analysed with a custom set of Interactive Data Language tools. The flat fields were generated from a median of dark subtracted exposures of the sky. Those flats were then normalised to a median value of one. The sky frames were generated as the median of the 15 dithered science frames; each science image was then sky-subtracted and flat-fielded. The reduced science frames were combined into a single combined image using an intra-pixel interpolation that conserves flux, shifts the individual dithered frames by the appropriate fractional pixels, and median-co-adds the frames. The resolution of the combined image was determined from the full-width half-maximum (FWHM) of the point spread functions: 0.105\arcsec.  
        
The sensitivity of the combined AO image to the companions was determined by injecting simulated sources around the primary target at separations of integer multiples of the central source's FWHM and every $20^\circ$ in azimuth \citep{furlan2017}. The brightness of each injected source was scaled until standard aperture photometry detected it with $5\sigma$ significance, and the brightness relative to TOI-3884 of the faintest injected sources that was detected set the contrast limit at that injection location. The adopted $5\sigma$ limit at each separation is then the average of the limits for the 18 azimuths at that radial distance, and the uncertainty on the limit is their root mean square dispersion. The resulting sensitivity curve for the Palomar image is shown in Fig.~\ref{fig:palomar_ao}; no stellar companions were detected.
    
\subsection{Gaia assessment}
We used Gaia to probe for wide stellar companions outside the field of view of the PHARO image that would be bound to TOI-3884. Such stars are typically already in the TESS Input Catalog, and their flux dilution to the transit has therefore already been accounted for in the transit fits and the associated derived parameters. Based upon parallax and proper motion similarity \citep[e.g.][]{mugrauer2020,mugrauer2021}, Gaia identifies no such companions.
    
The Gaia data release 3 (DR3) astrometry \citep{gaiaDR3} also contains indirect information on potential inner companions that would have gone undetected by both Gaia and the AO imaging. The Gaia renormalised unit weight error (RUWE) is a goodness of fit metric, akin to a reduced chi-square. Values ${\lesssim}~1.4$ indicate that the Gaia astrometry is consistent with a single star, whereas RUWE values ${\gtrsim}~1.4$ indicate excess astrometric noise, which can be from an unseen companion \citep[e.g.][]{ziegler2020}. TOI-3884 has a Gaia DR3 RUWE value of 1.25, and its Gaia astrometry is therefore fully consistent with a single star.   
    
\subsection{ESPRESSO}

We obtained RV measurements of TOI-3884 with ESPRESSO \citep{pepe2021}, a fibre-fed spectrograph installed at the Very Large Telescope at Paranal Observatory. Under the ESO programme 109.24EP, we collected two reconnaissance spectra with an exposure time of 1300~seconds, 2x1 binning, using the slow readout mode, and the single-telescope HR mode (R$\sim$140\,000). During the exposure we elected to maintain the calibration fibre on sky. The signal-to-noise ratio at 756~nm for the two measurements were 25 and 29, equivalent to expected RV precisions of 1.9 m/s and 1.6 m/s. The data reduction and RV extraction were done through the ESPRESSO-ESO pipeline\footnote{\url{https://www.eso.org/sci/software/pipelines/}}. The cross-correlation function (CCF) was derived by using the M4 mask available in the pipeline. The RVs and CCF characteristics are listed in Table~\ref{table.rv}. Visual inspection of the CCFs clearly exclude the case of a double-line spectroscopic binary. We do measure a slight but significant variation in the contrast and FWHM of the CCFs between the two epochs. This could be due to a change in the contribution from the starspot to the averaged temperature.

\section{Stellar parameters}\label{section.stellar_parameters}

We used empirical relations to derive the mass and radius of TOI-3884 from its absolute magnitude in the $K_s$-band, which is less affected by stellar spots and other activity phenomena than bluer bands. We used the zero-point corrected \citep{lindegren2021} Gaia parallax \citep{gaia2016, gaiaEDR3} to compute the distance and obtain an absolute magnitude of $M_{K_s}~=~7.056~\pm~0.017$. We then used the empirical relations of \citep{mann2019} and \citep{mann2015} to derive a mass of $M_{\star}=0.2813\pm0.0067$~\Msun and a radius of $R_{\star}=0.3043~\pm~0.0090$~\Rsun, which we adopt as our preferred values (Table~\ref{table.stellar_parameters}). We derived an alternative stellar radius from the spectral energy distribution (SED), which we constructed using photometry from Gaia \citep{riello2021}, the 2-Micron All-Sky Survey \citep[2MASS,][]{2mass,cutri2003}, and the Wide-field Infrared Survey Explorer \citep[WISE,][]{wise,cutri2013}. We modelled the photometry (Table~\ref{table.stellar_parameters}) using the Bayesian procedure described in \citet{diaz2014}, with the PHOENIX/BT-Settl \citep{allard2012} stellar atmosphere models. We used informative priors for the effective temperature ($T_{\mathrm{eff}} = 3269 \pm 70$~K, which corresponds to an M4 spectral type) and metallicity ($[\rm{Fe/H}] = 0.23 \pm 0.12$~dex), both derived from the co-added ESPRESSO spectra \citep[which we analysed with \specmatch;][]{yee2017}, and the distance from Gaia. We used non-informative priors for the rest of the parameters. We used one jitter term \citep{gregory2005} for each set of photometric bands (Gaia, 2MASS, and WISE). The parameters, priors, and posterior median, and the 68\% credible interval (CI) are listed in Table~\ref{table.sed}. The maximum a posteriori (MAP) model is shown in Fig.~\ref{fig.sed}. The derived radius ($R_{\star}=0.3005 \pm 0.0090$~\Rsun), is compatible with the adopted one. Evolution models predict that TOI-3884 is fully convective \citep{chabrier1997}. We repeated the SED modelling with non-informative priors for the effective temperature and metallicity, and obtained an effective temperature of 3270$^{+270}_{-170}$~K, compatible with the one derived with the ESPRESSO spectra, but less precise.

Rotational broadening can be measured with the FWHM of the CCF \citep{santos2002}. This usually requires calibrating the FWHM$_0$ for a set of stars with a range of temperatures and with unresolved stellar rotations. We did not perform such a full calibration but, instead, we picked one other M dwarf, LHS\,1140, with a similar spectral type (M4.5) to TOI-3884, a slow rotation (131~days), and also observed with ESPRESSO \citep{lillo-box2020}. We reduced LHS\,1140 spectra taken at three epochs (2019 December 9, 14, and 15) and measured an average CCF FWHM of 5.29 km/s. In comparison, we measured an FWHM = 5.46~km/s for TOI-3884. We then followed \citet{hirano2010} to derive \vsini\ as $(1.23\;\vsini)^2 = {\rm FWHM}_{\rm TOI-3884}^2 - {\rm FWHM}_{\rm 0,\;LHS\,1140}^2$, and obtained a value of $\vsini$ = 1.1~km/s. Given $R_\star$ from Table~\ref{table.stellar_parameters} and $i_\star$ from Table~\ref{table.spot}, this corresponds to a rotation period of $\sim$18 days. In addition, we measured $\log R'_{HK} = -4.79 \pm 0.13$ in the co-added ESPRESSO spectrum, from which the \cite{astudillo2017} activity-rotation relation estimates a rotation period of $P_{\rm rot} = 30 \pm 6$~days; this measurement must be taken with caution though, because of the poor signal-to-noise ratio of the spectrum at the H and K Ca~{\rm II} lines.

\section{Analysis}\label{section.analysis}

The shape of the asymmetric transit of TOI-3884~b seems stable over at least two years, and it changes with observing wavelength (chromatic effect). In addition, the TESS light curve shows no signs of rotational modulation. We explored two models that can explain such characteristics: a rotationally flattened and gravity darkened very rapidly rotating star \citep{barnes2009,dholakia2022}, and a polar spot. The rapidly rotating star model is strongly excluded by the ESPRESSO spectra because it needs \vsini\ of 350~km/s. Therefore, the polar spot is our adopted model. It should be noted that a circumpolar inhomogeneity such as a ring would match the data as well as a polar spot.

We first modelled the TESS, ExTrA, and LCOGT transits and the ESPRESSO RVs with \juliet \citep{espinoza2019}, which uses \batman \citep{kreidberg2015} for its transit model and \radvel \citep{fulton2018} to model the RVs. We added a Gaussian process (GP) regression model, with an approximate Matern kernel \citep[\texttt{celerite},][]{foreman-mackey2017} for the model of the error terms of the photometry (with different kernel hyperparameters for each transit observation, except for the TESS data, for which we used common kernel hyperparameters within each sector), to account for the asymmetry of the transit, and instrument-dependent dilution factors to deal with the chromaticity. We used a quadratic limb-darkening law \citep{manduca1977}, parameterised following \citet{kipping2013}. We used a stellar density prior from Sect.~\ref{section.stellar_parameters}, and adopted non-informative priors for the rest of the parameters. We sampled from the posterior with \dynesty \citep{speagle2020}. Table~\ref{table.spot} lists the parameters, priors, and posteriors. Figures~\ref{fig.juliet} and \ref{fig.RVplot} show the data and the model posterior. 

\begin{figure}
  \centering
  \includegraphics[width=0.48\textwidth]{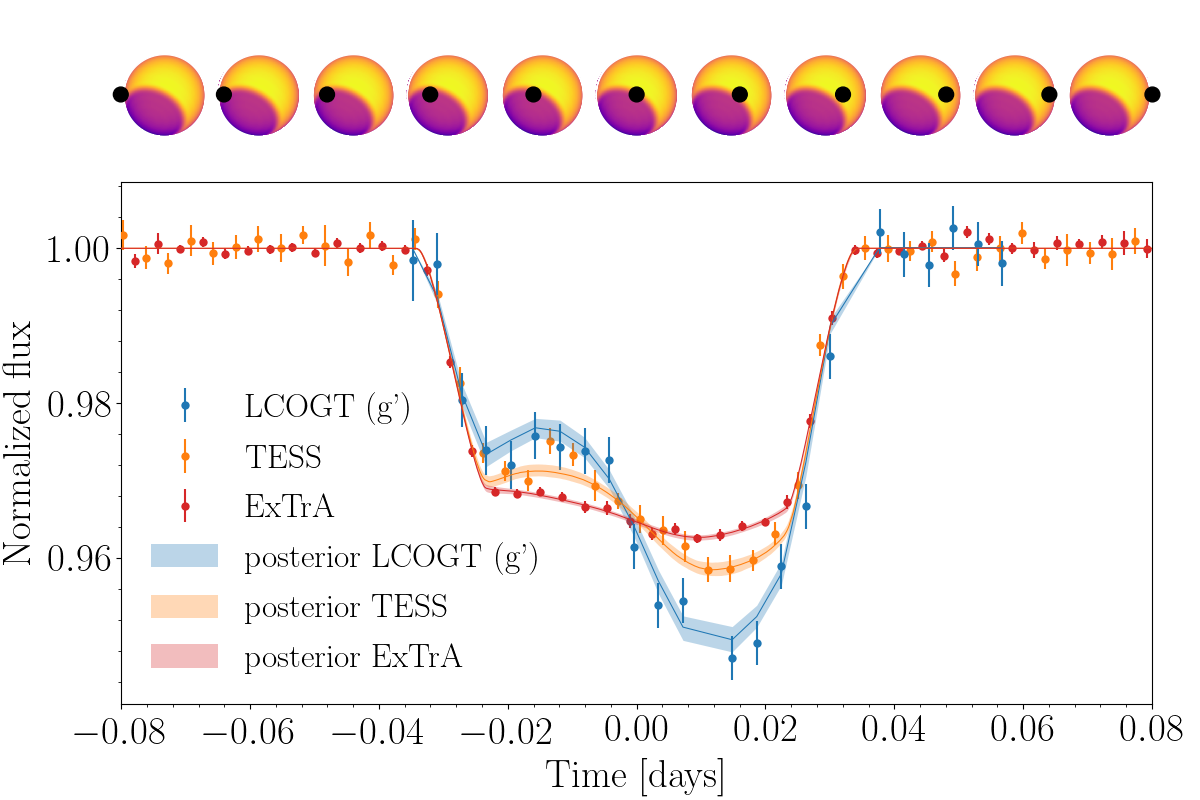}
  \caption{Phased transits observed by LCOGT, TESS (only sectors 46 and 49), and ExTrA, and the posterior model of a planet transiting a star with a polar spot. For clarity, the figure only shows 5-minute bins in the phased light curve for TESS and ExTrA. {\it Above the panel}: Model of the stellar surface and the planet for the median posterior values, as seen by TESS at different times during the transit.} \label{fig.starry_spot}
\end{figure}

We performed a second analysis, using only the transit data, to model the starspot. Within the uncertainties of the light curves, the shape of the transit remained constant at each wavelength (Fig.~\ref{fig.residuals}). We therefore phase-folded the light curves\footnote{We previously checked, using \juliet, that there are no significant transit timing variations (Fig.~\ref{fig.TTVs}).} to speed up the spot modelling computations. To do this, we used the MAP model of the analysis with \juliet to normalise the TESS\footnote{We did not use the sector 22 transits, observed with a 30-minute cadence, which blurs their shape.}, ExTrA, and LCOGT transits (imposing a straight line between the first and fourth contacts in order to not remove the transit or the asymmetry), and folded the transits according to the posterior ephemeris of the analysis with \juliet (Table~\ref{table.spot}). 
We modelled the folded transits in the three bands (g', TESS, and ExTrA) with the \starry code \citep{luger2019,luger2021a,luger2021b}. We describe the surface of the star with a quadratic limb-darkening law (as in the modelling with \juliet), and a top hat function in angular separation (from a longitude and latitude of the stellar surface) described with spherical harmonic expansion up to degree 30, representing a circular spot with uniform contrast. As we assume a polar spot, we fixed its latitude to 90\degree; thus, the model is independent of the value of its longitude and the rotation period of the star. The inclination of the spin axis and the sky-projected spin-orbit angle, which determine the position of the pole relative to the transit cord, are free parameters of the model. The spot contrast\footnote{The spot contrast is defined in \starry as the fractional change in the intensity at the centre of the spot ($I_{\rm spot}$) relative to the baseline intensity of an unspotted stellar surface ($I_{\rm photosphere}$): ${\rm contrast} = 1 - \frac{I_{\rm spot}}{I_{\rm photosphere}}$.} is a free parameter for each of the g', TESS, and ExTrA bands, which have a respective $\lambda_{\rm pivot}$ \citep{koornneef1986} of 476, 770, and 1165~nm. We oversampled the model in time and accounted for the integration time of the observations through binning \citep{kipping2010}. We used the \pymc package \citep{salvatier2016,hoffman2011} to sample from the posterior. The list of parameters, priors, and posteriors are shown in Table~\ref{table.spot}, and the data and model posterior are shown in Fig.~\ref{fig.starry_spot}. Figure~\ref{fig.spot_contrast} shows the spot contrast for the three bands. From these three spot contrasts, we estimate a photosphere-spot difference temperature of 187 $\pm$ 21~K (Sect.~\ref{sec.contrast}), which is in the range measured for M dwarfs \citep{berdyugina2005,barnes2015}.

Follow-up transit observations combined with the high-resolution imaging rule out most nearby eclipsing binary scenarios. Thanks to the significant proper motion of the target, historical images exclude background eclipsing binaries (Fig.~\ref{fig.DSS}). This leaves as a false positive scenario only a bound triple star, in which an eclipsing binary is diluted by a third star. The high-resolution imaging, the Gaia RUWE, and the single-lined ESPRESSO spectra excluded some of these configurations. As an additional check, we repeated the modelling with \starry, allowing for a different planet-to-star radius ratio in each band. The three ratios are then compatible (see Fig.~\ref{fig.RadiusRatio}), and reinforce the planetary scenario. The ESPRESSO RVs, the CCF contrast, and the CCF bisector span are compatible with a single star, and confirm the planetary nature of TOI-3884~b.

\section{Results and discussion}\label{section.results}

\begin{figure*}
  \centering
  \includegraphics[width=0.33\textwidth]{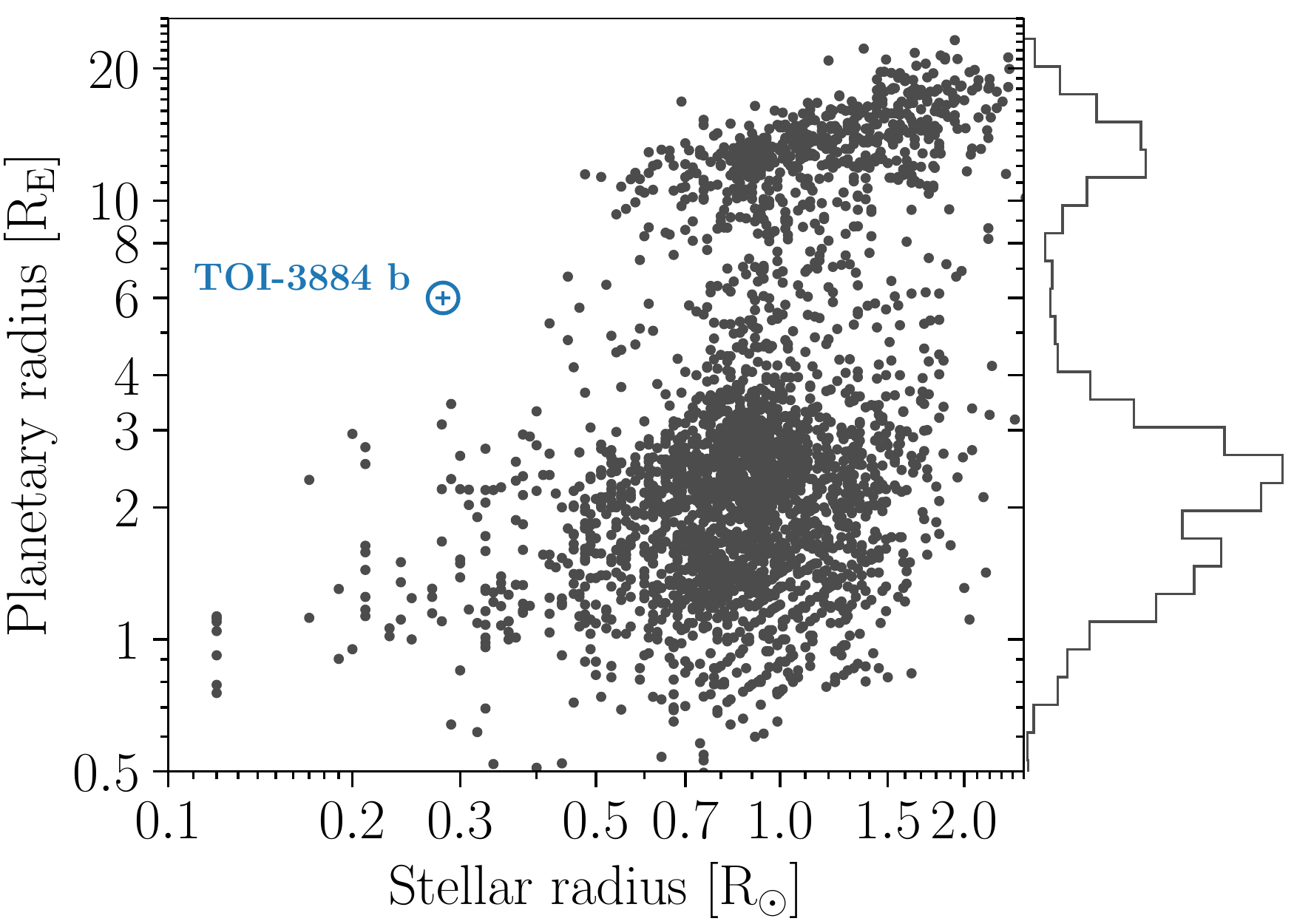}
  \includegraphics[width=0.33\textwidth]{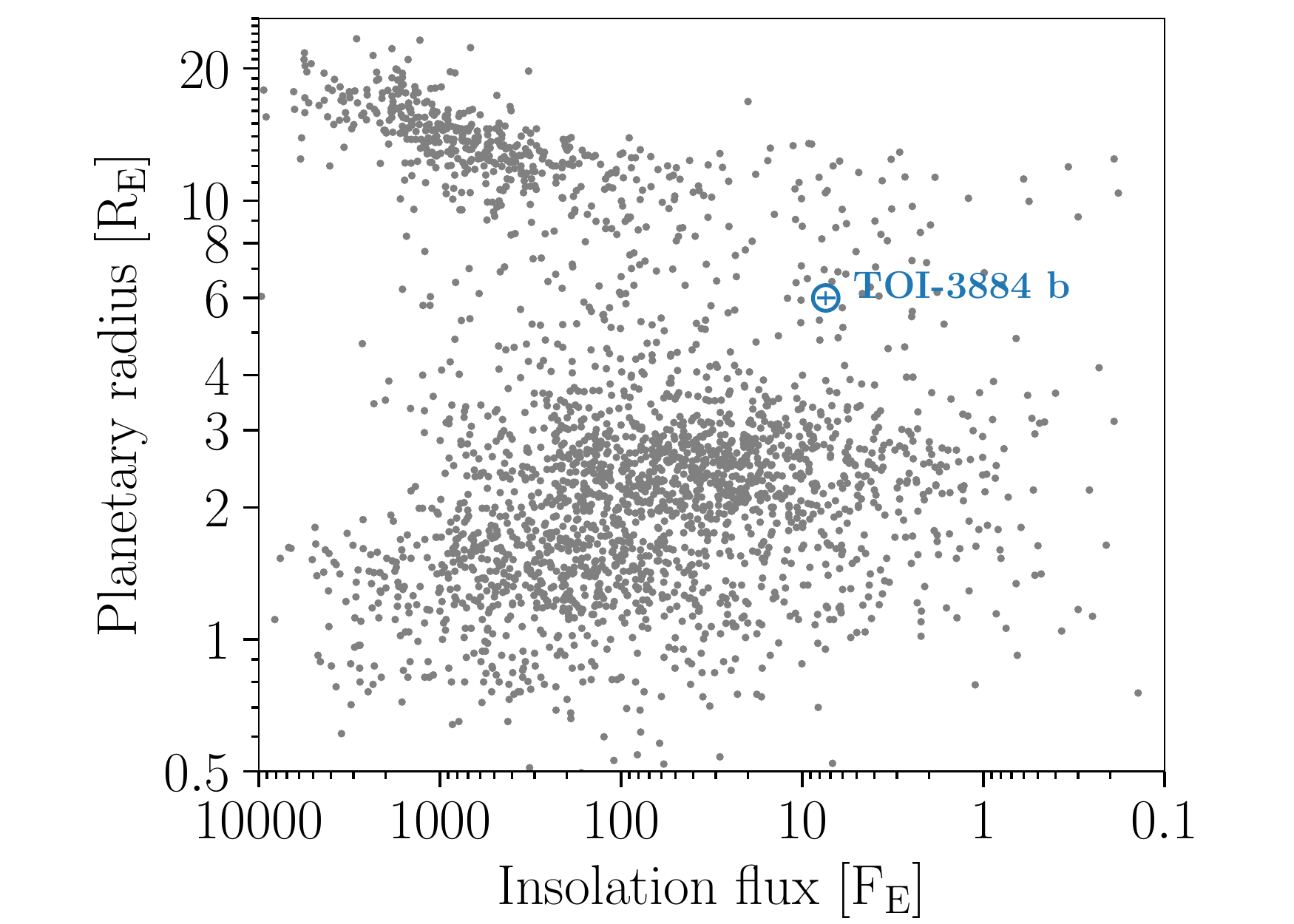}
  \includegraphics[width=0.33\textwidth]{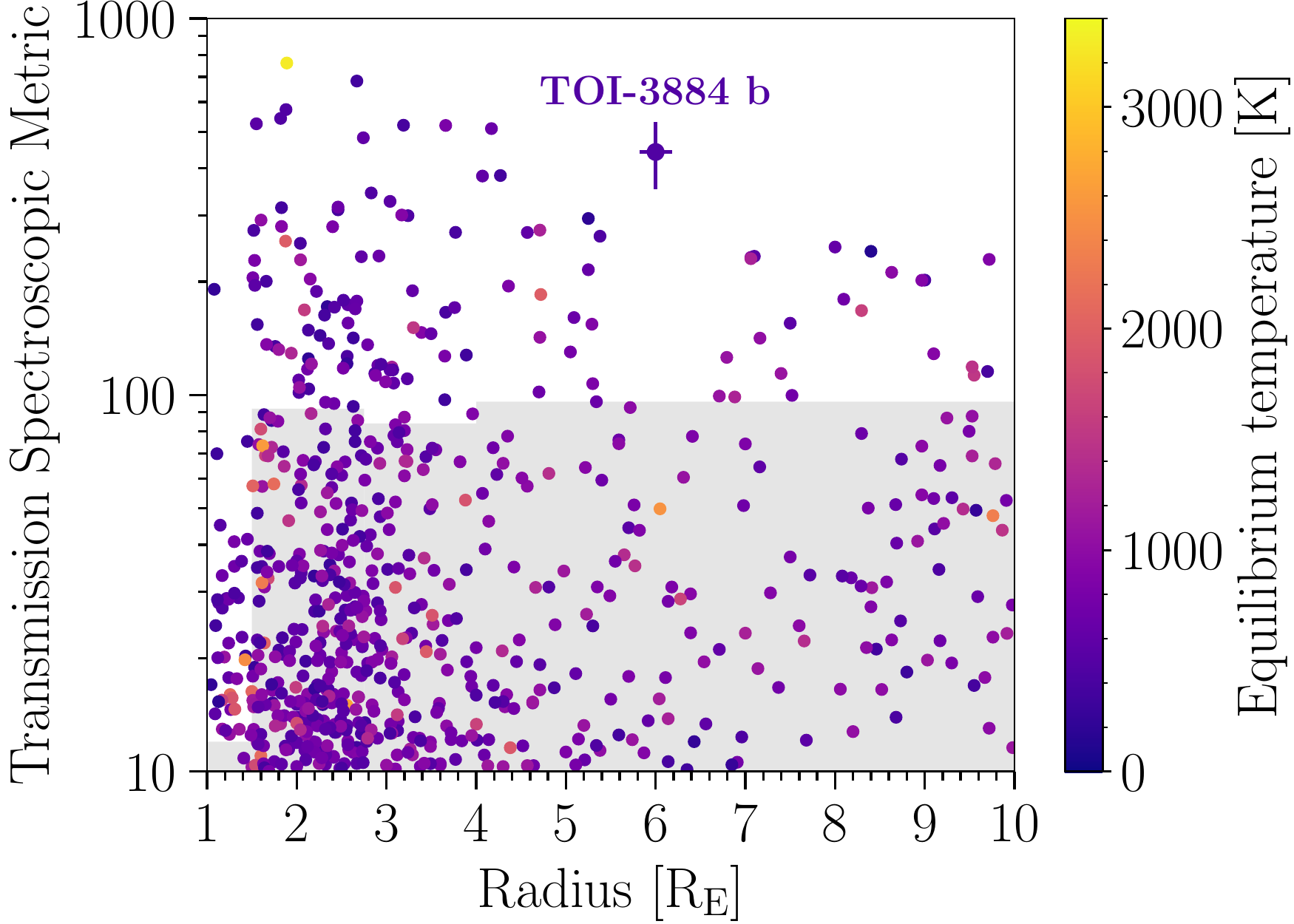}
  \caption{TOI-3884~b in context. {\it Left}: Planetary versus stellar radius. Black dots are transiting planets listed in the NASA Exoplanet Archive (\url{https://exoplanetarchive.ipac.caltech.edu/} with a planetary radius uncertainty smaller than 20\%. The blue circle with error bars marks the position of TOI-3884~b. The histogram of known planets is shown at the right of the panel. {\it Centre}: Planet radius versus insolation flux. Error bars mark the position of TOI-3884~b (blue) and grey dots are planets listed in the NASA Exoplanet Archive with a radius uncertainty smaller than 20\%. {\it Right}: Transmission spectroscopy metric for planets with a radius between 1 and 10~\REarth. Data are taken from the NASA Exoplanet Archive. Grey areas are below the cutoffs that \citet{kempton2018} suggested for follow-up efforts. The symbol with error bars is TOI-3884~b. The colour indicates the equilibrium temperature of the planet, computed for zero albedo and full day-night heat redistribution.} \label{fig.results}
\end{figure*}

The TOI-3884 system is composed of an M4 dwarf star orbited by a $6.00 \pm 0.18$~\REarth planet, TOI-3884~b, with a period of 4.5~days. This planet size, between gas Giants and Neptunians, is especially rare around low-mass stars (Fig.~\ref{fig.results}). The very preliminary mass that we obtain for TOI-3884~b, assuming neither stellar activity nor other planets contribute to the RVs, is consistent with that of Neptune, but the derived planet density would be one-fourth that of Neptune. The transmission spectroscopy metric \citep[TSM,][]{kempton2018} of TOI-3884~b is $442 \pm 90$. As shown by Fig.~\ref{fig.results}, this TSM is the most favourable to date for an $R_p\sim6$~\REarth planet, owing to the small size of the host star and its relative brightness.

Follow-up investigations for atmospheric characterisation will require good constraints on the mass of the planet, to 20\% precision or better. Indeed, transmission spectroscopy depends on the scale height of the atmosphere, which itself depends on the planet mass. A firm mass measurement with additional RVs is therefore required to support further follow-up observations with the James Webb Space Telescope \citep[JWST,][]{gardner2006}.

Our preferred model has the planet occulting a giant polar spot of 49\degree\ radius, covering about 17\% of the stellar surface. Due to the inclination of the stellar spin axis, a similar spot on the opposite pole would be hidden from our viewpoint, and thus would not affect the observed properties of TOI-3884~b. 
A giant polar spot suggests a large-scale magnetic field that is axisymmetric and poloidal. This is in agreement with spectropolarimetric surveys of M dwarfs that have shown that stars with mass ranging from 0.15 to 0.5~\Msun, and with short rotation periods, generate strong and long-lived magnetic fields featuring a significant large-scale poloidal component (\citealt{donati2006}, \citealt{morin2008}, and for a general picture, see Fig.~14 of \citealt{moutou2017}). This magnetic topology is very common for rotation periods of less than 10~days, but there are weak constraints on whether this behaviour extends to longer periods of 20 days. CE Boo (GJ 569A; 0.45~\Msun, $P_{\rm rot} \sim 15$~d) is one of the few stars in this period range, close to that of TOI-3884, with a reconstructed magnetic map \citep{donati2008}, and its topology is almost completely poloidal and mostly axisymmetric. The magnetic activity of the star could be further studied with spectropolarimetry in the near infrared, for example with SPIRou \citep{donati2020}. This could be important because of the influence this large spot can have on the planet transmission spectroscopy.

The expected amplitude of the Rossiter–McLaughlin effect \citep{rossiter1924,mcLaughlin1924} for TOI-3884~b is 40 m/s, and this observation will provide an independent measurement of the obliquity that can validate the results from the polar spot model. A more detailed analysis of spectroscopic observations during transit can map the stellar surface occulted by the planet \citep{bourrier2021} and will test the presence of the polar spot. If confirmed, the polar spot allows the measurement of the true spin–orbit angle ($\psi$), and we find that it would be misaligned ($\psi = 50\pm12\degree$ or $130\pm12\degree$). This measurement will probe the history of the planet, which almost certainly had to migrate to end up so close to its star.

\begin{acknowledgements}
Funding for the TESS mission is provided by NASA's Science Mission Directorate. We acknowledge the use of public TESS data from pipelines at the TESS Science Office and at the TESS Science Processing Operations Center. This research has made use of the Exoplanet Follow-up Observation Program website, which is operated by the California Institute of Technology, under contract with the National Aeronautics and Space Administration under the Exoplanet Exploration Program. Resources supporting this work were provided by the NASA High-End Computing (HEC) Program through the NASA Advanced Supercomputing (NAS) Division at Ames Research Center for the production of the SPOC data products. This paper includes data collected by the TESS mission that are publicly available from the Mikulski Archive for Space Telescopes (MAST). This research has made use of the Exoplanet Follow-up Observation Program (ExoFOP; DOI: 10.26134/ExoFOP5) website, which is operated by the California Institute of Technology, under contract with the National Aeronautics and Space Administration under the Exoplanet Exploration Program.

We thank the ESO Director General for allocating discretionary observing time with ESPRESSO for reconnaissance spectroscopy of TOI-3884.

We are grateful to the ESO/La Silla staff for their support of ExTrA.

This work has made use of data from the European Space Agency (ESA) mission {\it Gaia} (\url{https://www.cosmos.esa.int/gaia}), processed by the {\it Gaia} Data Processing and Analysis Consortium (DPAC, \url{https://www.cosmos.esa.int/web/gaia/dpac/consortium}). Funding for the DPAC has been provided by national institutions, in particular the institutions participating in the {\it Gaia} Multilateral Agreement.

This work makes use of observations from the LCOGT network. Part of the LCOGT telescope time was granted by NOIRLab through the Mid-Scale Innovations Program (MSIP). MSIP is funded by NSF.
   
This research has made use of the Spanish Virtual Observatory (https://svo.cab.inta-csic.es) project funded by MCIN/AEI/10.13039/501100011033/ through grant PID2020-112949GB-I00.

The Pan-STARRS1 Surveys (PS1) and the PS1 public science archive have been made possible through contributions by the Institute for Astronomy, the University of Hawaii, the Pan-STARRS Project Office, the Max-Planck Society and its participating institutes, the Max Planck Institute for Astronomy, Heidelberg and the Max Planck Institute for Extraterrestrial Physics, Garching, The Johns Hopkins University, Durham University, the University of Edinburgh, the Queen's University Belfast, the Harvard-Smithsonian Center for Astrophysics, the Las Cumbres Observatory Global Telescope Network Incorporated, the National Central University of Taiwan, the Space Telescope Science Institute, the National Aeronautics and Space Administration under Grant No. NNX08AR22G issued through the Planetary Science Division of the NASA Science Mission Directorate, the National Science Foundation Grant No. AST-1238877, the University of Maryland, Eotvos Lorand University (ELTE), the Los Alamos National Laboratory, and the Gordon and Betty Moore Foundation.

Based on observations collected at the European Southern Observatory under ESO programme 109.24EP.

This work has been supported by a grant from Labex OSUG$@$2020 (Investissements d'avenir -- ANR10 LABX56).

Based on data collected under the ExTrA project at the ESO La Silla Paranal Observatory. ExTrA is a project of Institut de Plan\'etologie et d'Astrophysique de Grenoble (IPAG/CNRS/UGA), funded by the European Research Council under the ERC Grant Agreement n. 337591-ExTrA.

\end{acknowledgements}

\bibliographystyle{aa}
\bibliography{TOI-3884}

\begin{appendix}

\FloatBarrier
\section{Additional figures and tables}

\begin{figure*}
    \sidecaption
  \includegraphics[width=0.35\textwidth]{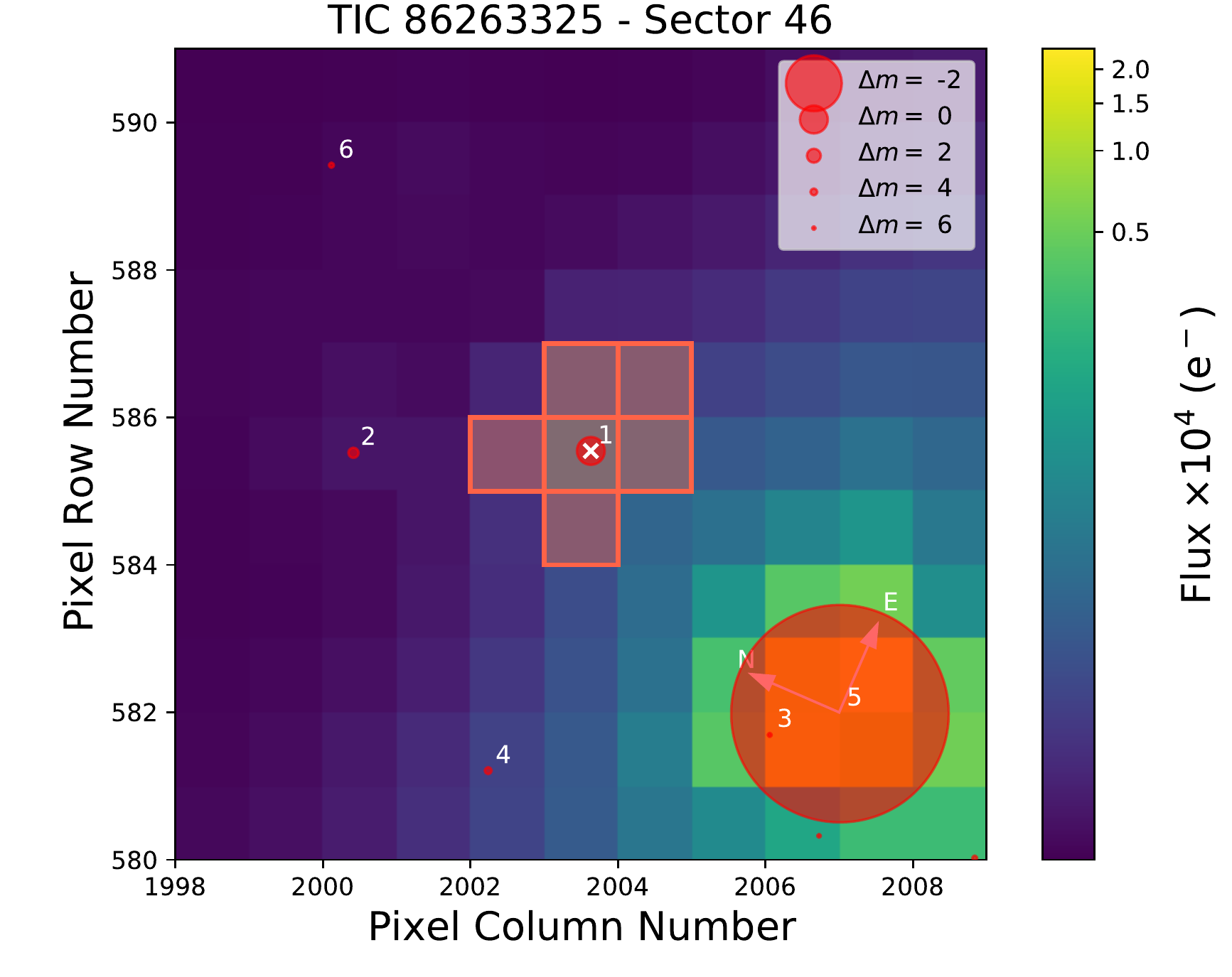}
  \includegraphics[width=0.35\textwidth]{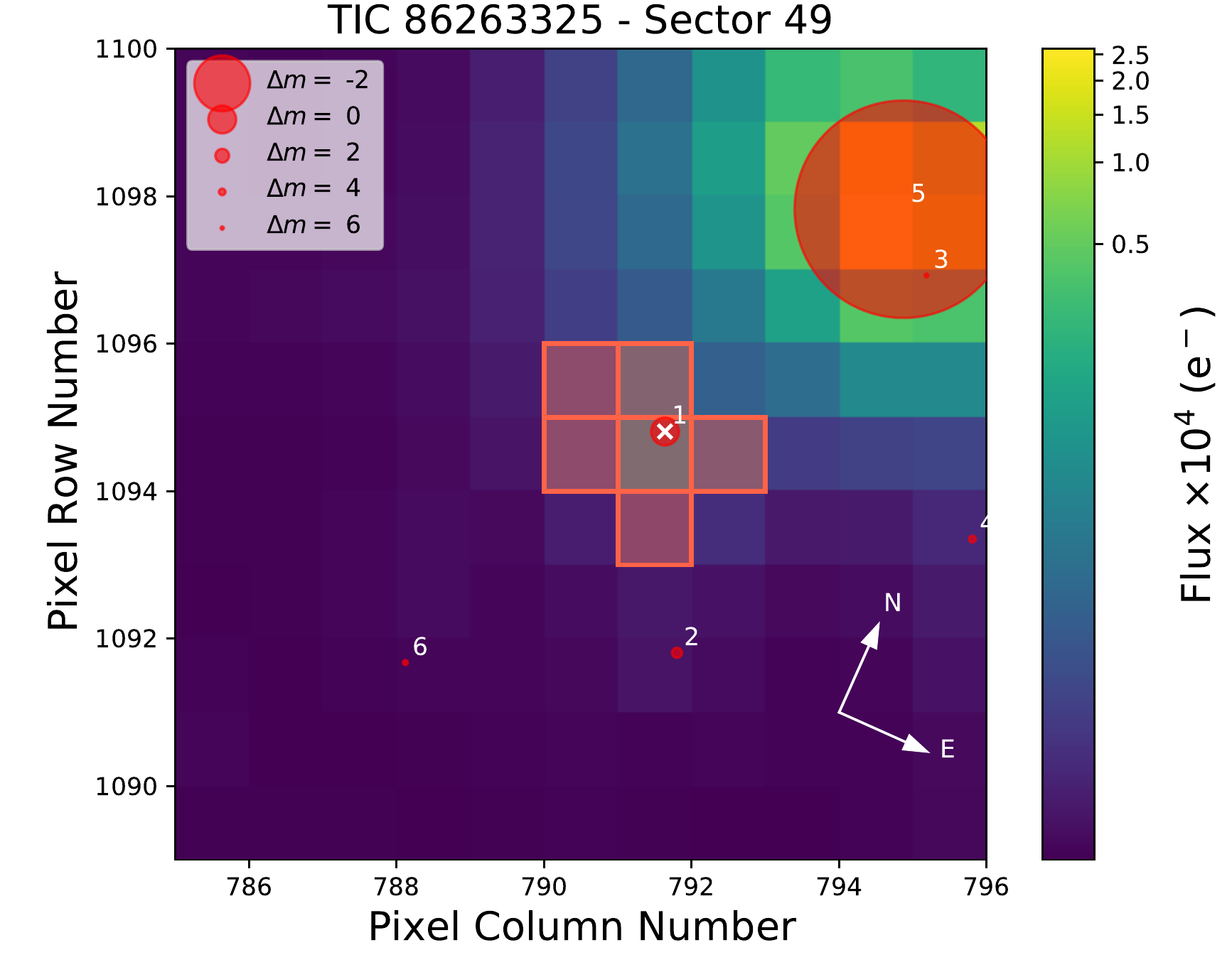}
  \caption{TESS target pixel file image of TOI-3884 in sectors 46 and 49 \citep[created with {\sc \tt tpfplotter},][]{aller2020}. The electron counts are colour-coded. The pixels highlighted in red are used for the simple aperture photometry. The positions of the stars in the Gaia DR2 \citep{gaia2018} are indicated with red circles (and labelled with numbers according to the distance to the main target in the aperture labelled `1'), and their sizes are proportional to the Gaia DR2 magnitudes.} \label{fig.tpfplotter}
\end{figure*}

\begin{figure}
  \centering
  \includegraphics[width=0.49\textwidth]{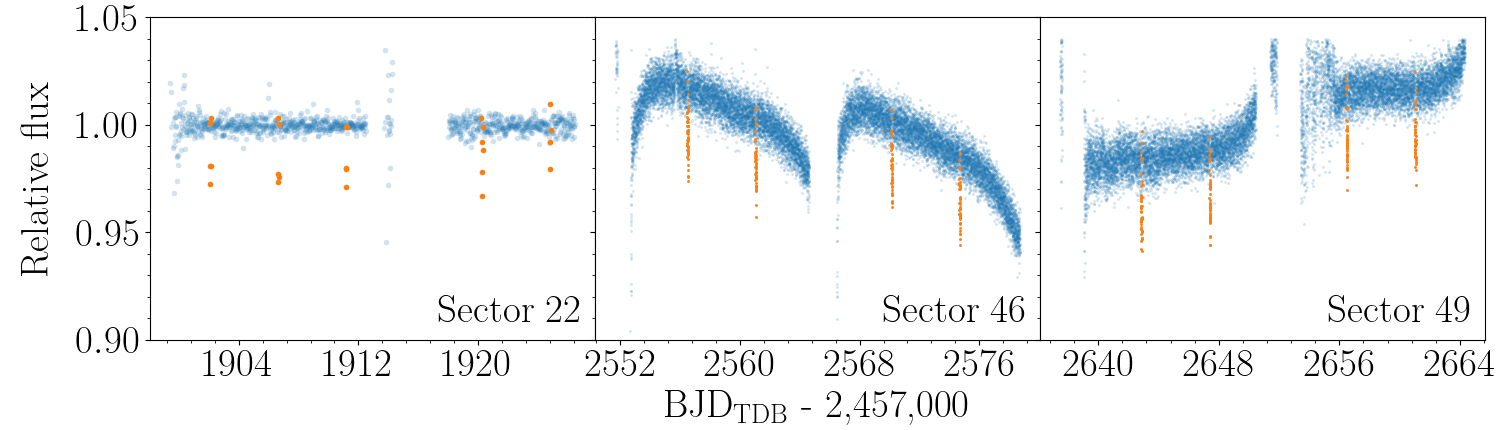}
  
  \includegraphics[width=0.49\textwidth]{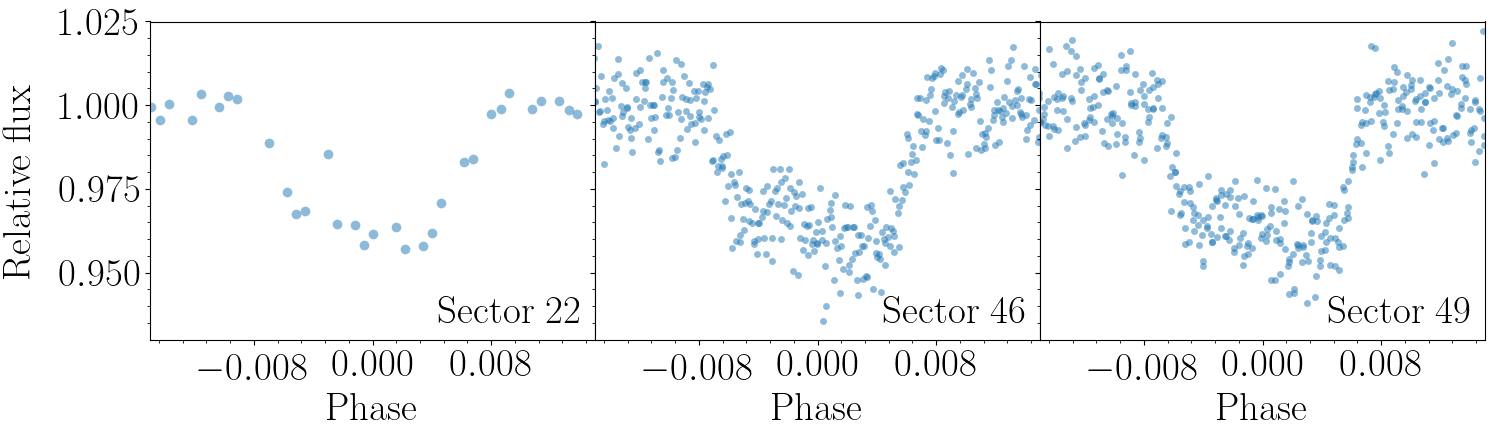}
  \caption{TESS observations. {\it Top}: Light curves from QLP KSPSAP (sector 22) and SPOC SAP (sectors 46 and 49) are shown by blue dots, with the transits of TOI-3884~b highlighted in orange. The SAP light curves do not show significant variability, apart from the one due to the satellite orbit. {\it Bottom}: Transits observed by TESS folded for each sector. } \label{fig.tess}
\end{figure}

\begin{figure}
    \centering
    \includegraphics[width=0.49\textwidth]{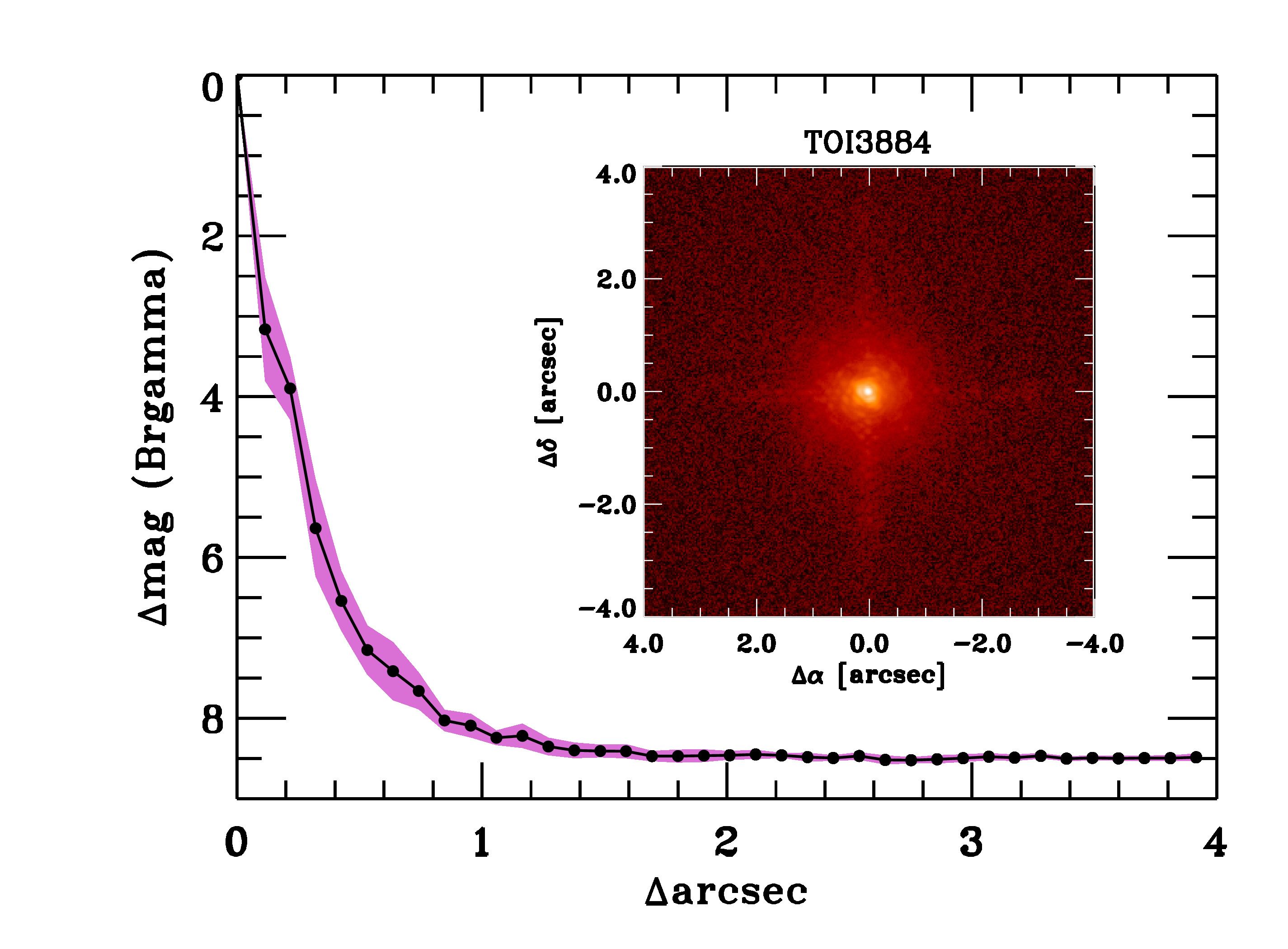}
    \caption{Palomar NIR AO imaging and sensitivity curves for TOI-3884 taken in the Br-$\gamma$ filter. The image reaches a contrast of $\sim 7$ magnitudes fainter than the host star within 0.5\arcsec. {\it Inset}: Image of the central portion of the data, centred on the star.
    }
    \label{fig:palomar_ao}
\end{figure}

\begin{table}[htb]
\setlength{\tabcolsep}{3pt}
\scriptsize
  \caption{Parameters measured on ESPRESSO spectra of TOI-3884.}\label{table.rv}
\begin{tabular}{lcccc}
\hline
\hline
 & &  CCF & CCF & CCF \\
Time & RV  & FWHM & contrast & bisector span \\
$[{\rm BJD_{TDB}}]$ & [km\,s$^{-1}$] & [km\,s$^{-1}$] & [\%] & [km\,s$^{-1}$] \\
\hline
2459762.484185 & $8.3318 \pm 0.0019$ & $5.4367 \pm 0.0037$ & $39.364 \pm 0.027$ & $0.0225 \pm 0.0037$ \\
2459773.488742 & $8.3589 \pm 0.0016$ & $5.4813 \pm 0.0031$ & $37.378 \pm 0.021$ & $0.0216 \pm 0.0031$ \\
\hline
\end{tabular}
\end{table}

\begin{table}[htb]
\tiny
  \caption{Astrometry, photometry, and stellar parameters for TOI-3884.}\label{table.stellar_parameters}
  \centering
    \setlength{\tabcolsep}{2.5pt}
\begin{tabular}{lcc}
\hline
\hline
    Parameter & Value & Source \\
     \hline
    Designations & LSPM J1206+1230 & \citet{lepine2005} \\
                 & TIC 86263325 & \citet{stassun2019} \medskip\\
    RA (ICRS, J2000)          & 12$^{\rm h}$ 06$^{\rm m}$ 17.44$^{\rm s}$ & Gaia EDR3 \\
    Dec (ICRS, J2000)         & $+12^{\rm o}$ 30' 24.9''                  & Gaia EDR3 \\
    $\mu$ RA [mas yr$^{-1}$]  & -186.042 $\pm$ 0.028                      & Gaia EDR3 \\
    $\mu$ Dec [mas yr$^{-1}$] & 26.388 $\pm$ 0.017                        & Gaia EDR3 \\
    Parallax [mas]            & 23.074 $\pm$ 0.026                        & Gaia EDR3$^a$ \\
    Distance [pc]             & 43.338 $\pm$ 0.049                        & Parallax \medskip\\

    Gaia-BP [mag] & 15.9710 $\pm$ 0.0047  & Gaia EDR3 \\
    Gaia-G [mag]         & 14.2463 $\pm$ 0.0029  & Gaia EDR3 \\
    Gaia-RP [mag]        & 12.9897 $\pm$ 0.0033  & Gaia EDR3 \\
    2MASS-J [mag]        & 11.127  $\pm$ 0.021  & 2MASS\\
    2MASS-H [mag]        & 10.552  $\pm$ 0.020 & 2MASS \\ 
    2MASS-Ks [mag] & 10.240  $\pm$ 0.017 & 2MASS \\
    WISE-W1 [mag]        & 10.157  $\pm$ 0.023 & WISE \\
    WISE-W2 [mag]        &  9.986  $\pm$ 0.021 & WISE \\
    WISE-W3 [mag]  &  9.762  $\pm$ 0.048 & WISE \medskip\\
    
    Mass, $M_\star$ [\Msun]                              & $0.2813 \pm 0.0067$          & \citet{mann2019}\\
    Radius, $R_\star$ [\Rsun]                            & $0.3043 \pm 0.0090$          & \citet{mann2015}\\
    Mean density, $\rho_{\star}$ [$\mathrm{g\;cm^{-3}}$] & $14.1^{+1.4}_{-1.2}$         & $M_\star$, $R_\star$\\
    Surface gravity, \logg\ [cgs]                        & $4.921 \pm 0.028$            & $M_\star$, $R_\star$ \smallskip\\
    Effective temperature, $T_{\rm eff}$ [K] & $3269 \pm 70$ & \specmatch \\
    Metallicity, [Fe/H] [dex] & $0.23 \pm 0.12$ & \specmatch \\
\hline
\end{tabular}
\tablefoot{$^a$ Corrected with \citet{lindegren2021}. $\mu$ is the proper motion.}
\end{table}

\begin{table}[htb]
\tiny
    \renewcommand{\arraystretch}{1.25}
    \setlength{\tabcolsep}{2pt}
\centering
\caption{Modelling of the SED.}\label{table.sed}
\begin{tabular}{lccc}
\hline
\hline
Parameter & & Prior & Posterior median   \\
&  & & and 68.3\% CI \\
\hline
Effective temperature, $T_{\mathrm{eff}}$ & [K]     & $N$(3269, 70)   & 3271$\pm$60 \\
Surface gravity, \logg\                   & [cgs]   & $U$(-0.5, 6.0)    & 5.24$^{+0.56}_{-0.83}$ \\
Metallicity, $[\rm{Fe/H}]$                & [dex]   & $N$(0.23, 0.12)    & 0.22$\pm$0.12 \\
Distance                                  & [pc]    & $N$(43.338, 0.049)& 43.338$\pm$0.048 \\
$E_{\mathrm{(B-V)}}$                      & [mag]   & $U$(0, 3)         & 0.071$^{+0.076}_{-0.049}$ \\
Jitter Gaia                               & [mag]   & $U$(0, 1)         & 0.181$^{+0.20}_{-0.083}$ \\
Jitter 2MASS                              & [mag]   & $U$(0, 1)         & 0.082$^{+0.099}_{-0.042}$ \\
Jitter WISE                               & [mag]   & $U$(0, 1)         & 0.046$^{+0.088}_{-0.034}$ \\
Radius, $R_\star$                         & [\Rsun] & $U$(0, 100)       & 0.3005$\pm$0.0090 \\
Luminosity                                & [L$_\odot$] &               & 0.00930$^{+0.00055}_{-0.00038}$ \smallskip\\
\hline
\end{tabular}
\tablefoot{$N$($\mu$, $\sigma$): Normal distribution prior with mean $\mu$, and standard deviation $\sigma$. $U$(l, u): Uniform distribution prior in the range [l, u].}
\end{table}

\begin{figure}
  \centering
  \includegraphics[width=0.49\textwidth]{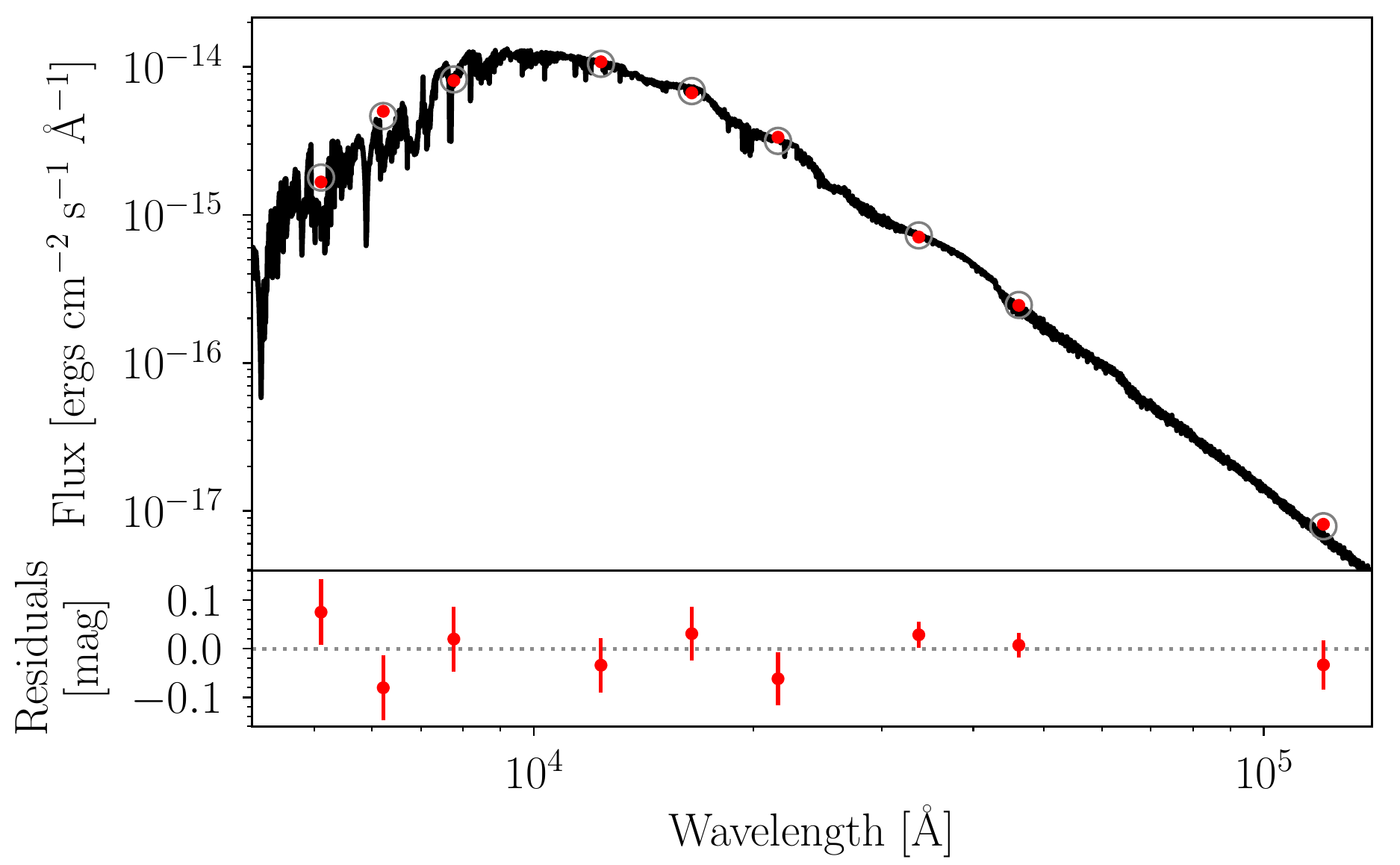}
  \caption{Spectral energy distribution of TOI-3884. {\it Top panel}: Maximum a posteriori PHOENIX/BT-Settl interpolated synthetic spectrum (solid line). The red circles show photometric observations and the grey open circles are the result of integrating the synthetic spectrum in the observed bandpasses. {\it Bottom panel}: Residuals of the MAP model (the jitter has been added quadratically to the error bars of the data).} \label{fig.sed}
\end{figure}

\begin{figure}
  \centering
  \includegraphics[width=0.49\textwidth]{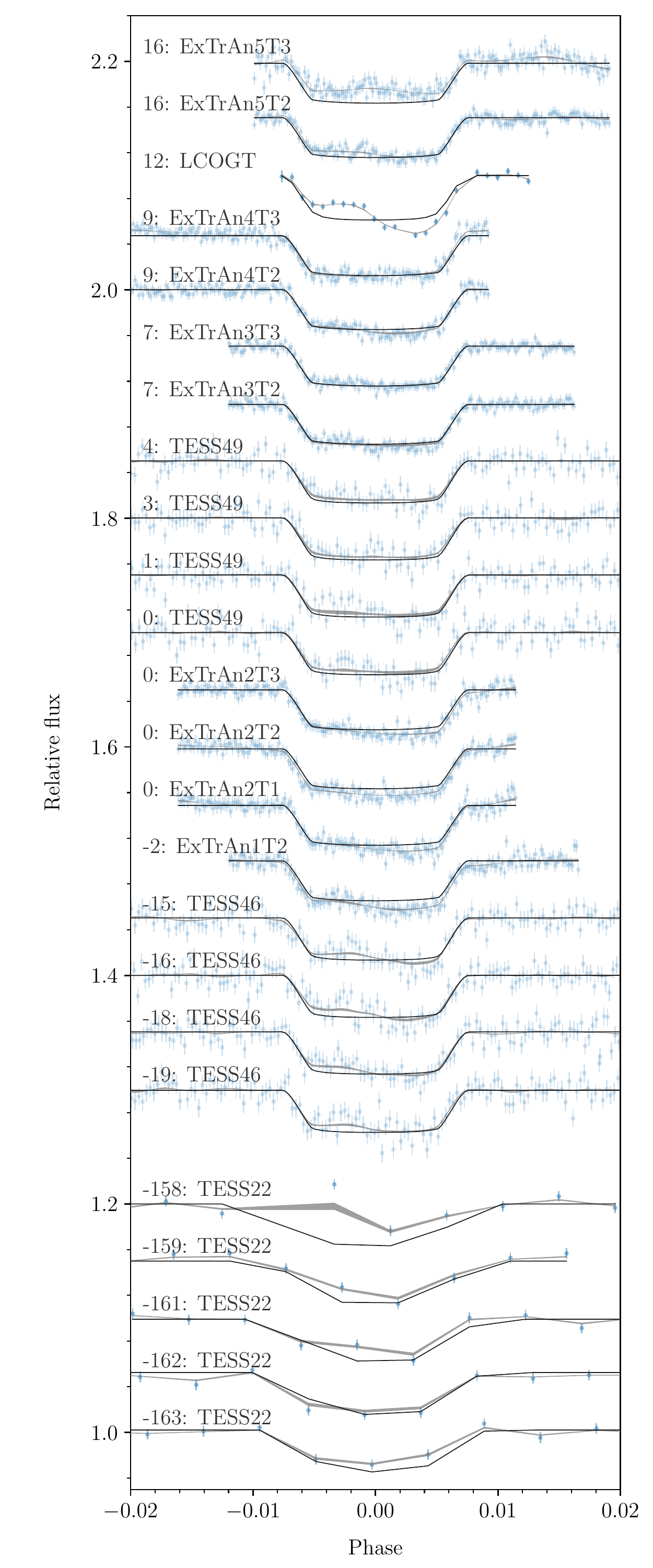}
  \caption{TESS, ExTrA, and LCOGT transits (offset for clarity) modelled with \juliet. The blue symbols with error bars are the data, the grey line is the \juliet posterior median model, and the black line is the corresponding pure transit model. Each transit is labelled with the epoch relative to T$_0$, the sector for transits observed with TESS, and the night and the telescope for transits observed with ExTrA.} \label{fig.juliet}
\end{figure}

\begin{figure}
  \centering
  \includegraphics[width=0.49\textwidth]{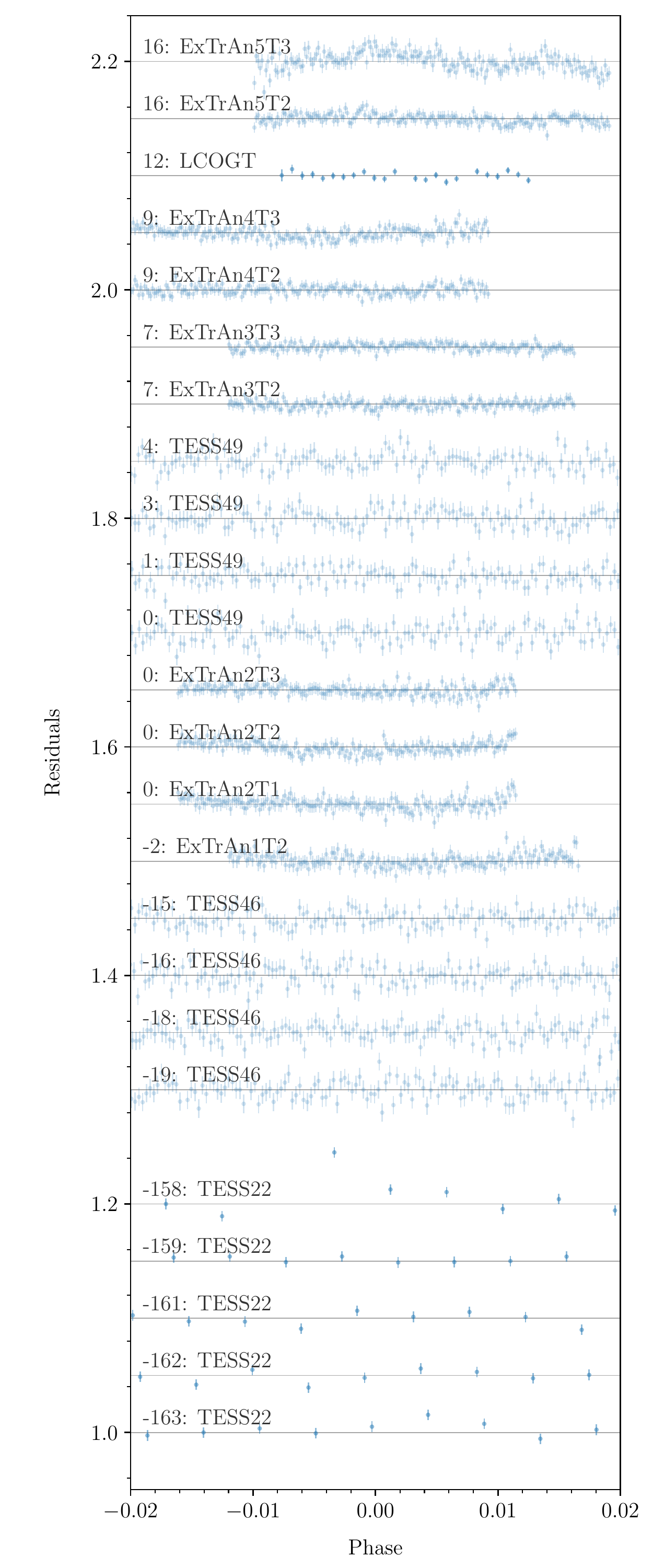}
  \caption{Transit observation (Fig.~\ref{fig.juliet}) residuals (data divided by model) from the posterior median model of a planet transiting a star with a polar spot from the analysis with the \starry code.} \label{fig.residuals}
\end{figure}

\begin{figure}
  \includegraphics[width=0.47\textwidth]{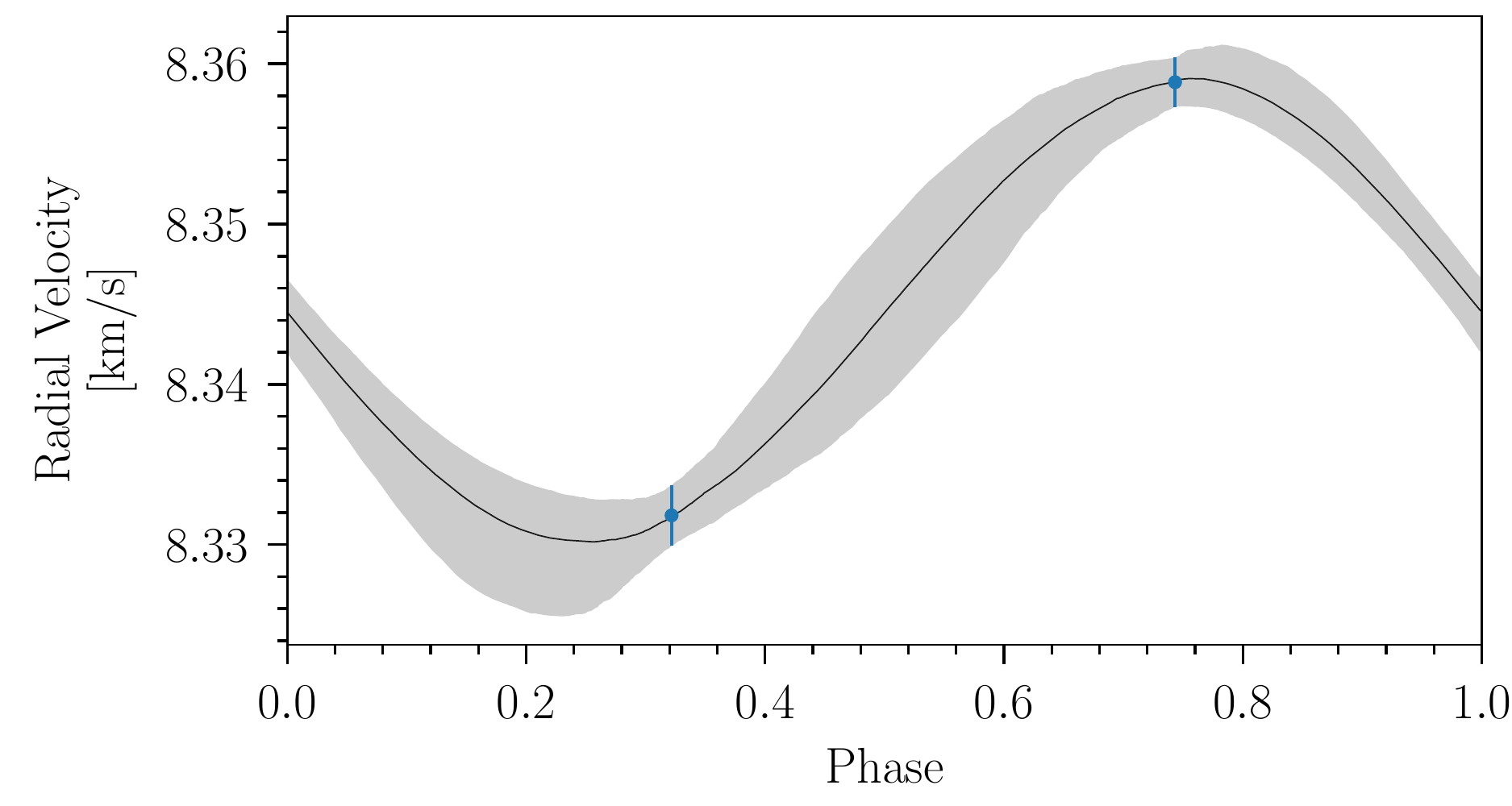}
  \caption{Phased ESPRESSO RVs (blue error bars, as computed from the spectra), median posterior model (black line), and 68\% CI (grey band).} \label{fig.RVplot}
\end{figure}

\begin{figure}
  \centering
  \includegraphics[width=0.49\textwidth]{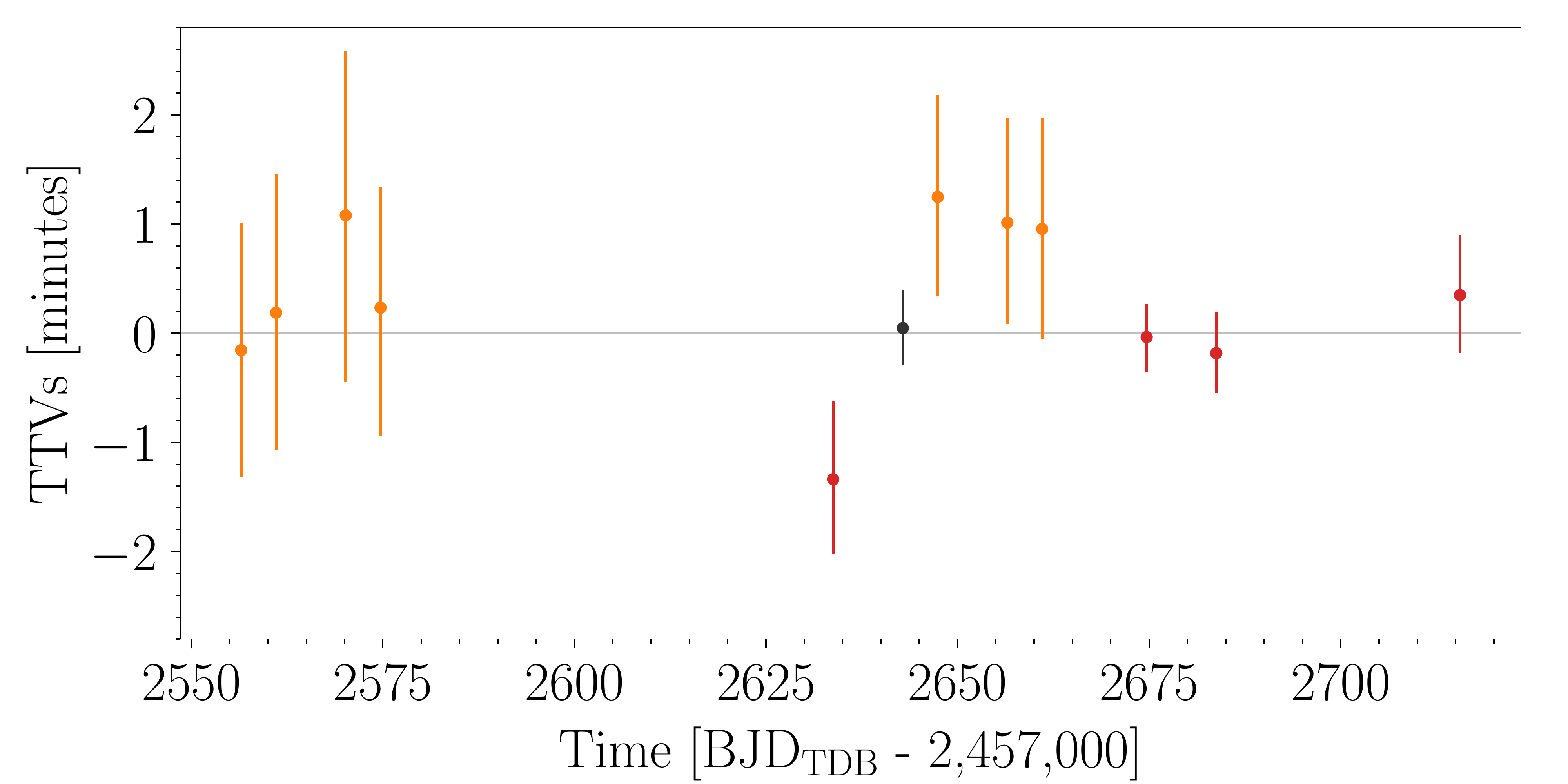}
  \caption{Transit timing variations from TESS sectors 46 and 49 (orange), ExTrA (red), and a transit observed simultaneously by TESS and ExTrA (black).} \label{fig.TTVs}
\end{figure}

\begin{table*}
  \tiny
  \setlength{\tabcolsep}{5pt}
\renewcommand{\arraystretch}{1.25}
\centering
\caption{Inferred system parameters.}\label{table.spot}
\begin{tabular}{lcccccc}
\hline
\hline
Parameter & Units & Prior &  Median and 68.3\% CI & Prior &  Median and 68.3\% CI & Adopted \\
 &  & \juliet &  \juliet & \starry & \starry & value \\
\hline
\emph{\bf Star (TOI-3884)} \\
Mean density, $\rho_{\star}$     & [$\mathrm{g\;cm^{-3}}$] & $N(14.1, 1.4)$ & $14.3 \pm 1.1$ & $N(14.1, 1.4)$ & $13.83^{+0.66}_{-0.99}$     & \juliet \\
Inclination of the spin axis, $i_\star$ & [\degree]        &                               & & $U(0, 90)$      & $47.0 \pm 8.5$      &  \\
Sky-projected spin-orbit angle, $\lambda$& [\degree]              &                               & & $U(90, 270)$    & $151 \pm 11$     &  \\
Polar spot size, $\alpha$        & [\degree]               & &                               & $U(0, 90)$      & $48.6 \pm 4.7$            &  \\
Polar spot contrast g'           &                         & &                               & $U(0, 0.9)$     & $0.664 \pm 0.097$           &  \\
Polar spot contrast TESS         &                         & &                               & $U(0, 0.9)$     & $0.365^{+0.064}_{-0.039}$  &  \\
Polar spot contrast ExTrA        &                         & &                               & $U(0, 0.9)$     & $0.159 \pm 0.025$        &  \\
$q_1$ g' &                       & $U(0, 1)$ & $0.27^{+0.37}_{-0.20}$                        & $U(0, 1)$       & $0.51 \pm 0.25$          & \starry \\
$q_2$ g' &                       & $U(0, 1)$ & $0.36^{+0.37}_{-0.26}$                        & $U(0, 1)$       & $0.29^{+0.33}_{-0.21}$   & \starry \\
$q_1$ TESS &                     & $U(0, 1)$ & $0.132^{+0.12}_{-0.077}$                      & $U(0, 1)$       & $0.224 \pm 0.13$          & \starry \\
$q_2$ TESS &                     & $U(0, 1)$ & $0.105^{+0.19}_{-0.081}$                      & $U(0, 1)$       & $0.122^{+0.19}_{-0.097}$ & \starry \\
$q_1$ ExTrA  &                   & $U(0, 1)$ & $0.027^{+0.028}_{-0.014}$                     & $U(0, 1)$       & $0.069^{+0.050}_{-0.027}$ & \starry \\
$q_2$ ExTrA  &                   & $U(0, 1)$ & $0.54 \pm 0.31$                               & $U(0, 1)$       & $0.61 \pm 0.25$           & \starry \\
Systemic velocity, $\gamma$      &[\kms]    & $U$(8.31, 8.37) & $8.3444 \pm 0.0015$ &   &  & \\
\smallskip\\

\emph{\bf Planet (TOI-3884~b)} \\
Semi-major axis, $a$                   & [au]             & & $0.0354 \pm 0.0014$ &                 &          &  \\
Eccentricity, $e$                      &                  & & < 0.32$^\dagger$, ($0.059^{+0.14}_{-0.044}$)  & $U(0.0, 0.9)$ & < 0.19$^\dagger$, ($0.024^{+0.063}_{-0.018}$) &  \juliet \\
Argument of pericentre, $\omega$       & [\degree]        & & $190 \pm 150$ & $U(0, 360)$ & $72^{+150}_{-62}$ & \juliet \\
Inclination, $i_p$                     & [\degree]        & & $89.83^{+0.12}_{-0.16}$ & $U(88, 92)$ & $90.10 \pm 0.29$   & \starry \\
Radius ratio, $R_{\mathrm{p}}/R_\star$ &                  & $U(0.1, 0.3)$ & $0.1899 \pm 0.0064$ & $U(0.1, 0.3)$   & $0.1808 \pm 0.0014$         & \starry \\
Scaled semi-major axis, $a/R_{\star}$  &                  & & $25.01 \pm 0.65$ &                 &             &  \\
Impact parameter, $b$                  &                  & $U(0, 1)$ & $0.072 \pm 0.072$ &                 & $-0.04 \pm 0.13$           &  \starry \\ 
Transit duration, $T_{14}$             & [h]              & & $1.646 \pm 0.011$ &                 &  &  \\
T$_0$ \;-\;2\;450\;000 & [BJD$_{\mathrm{TDB}}$]   & $U(9642.8624, 9642.8638)$ & $9642.86314 \pm 0.00012$    & & & \\
Orbital period, $P$                      & [d]              & $U(4.54447, 4.54468)$     & $4.5445697 \pm 0.0000094$   & & &  \\
RV semi-amplitude, K             &[\ms]     & $U$(0, 100)                        & $14.9^{+3.4}_{-1.6}$ &  & &  \\

Radius, $R_{\mathrm{p}}$               &[\Renom]          & & $6.31 \pm 0.28$ &                 & $6.00 \pm 0.18$             &  \starry \\
Mass, $M_{\mathrm{p}}$                 &[\MEarth]          & & $16.5^{+3.5}_{-1.8}$ &                 &              & \ \\
Mean density, $\rho_{\mathrm{p}}$ &[$\mathrm{g\;cm^{-3}}$]& & $0.365^{+0.095}_{-0.061}$ &            & $0.424^{+0.10}_{-0.061}$ & \starry \\
Surface gravity, $\log$\,$g_{\mathrm{p}}$ &[cgs]          & & $2.612 \pm 0.090$ &            & $2.656^{+0.088}_{-0.059}$ & \starry \\
Equilibrium temperature, T$_{\rm eq}$  & [K]              & & $463 \pm 12$ &                 &                 &   \\
Insolation flux                        & [F$_{\rm E}$]    & & $7.42 \pm 0.77$&                 &                & \smallskip\\

True spin–orbit angle, $\psi$         & [\degree]         & & &                 & $50 \pm 12$ or $130 \pm 12$ & \smallskip\\

$\sqrt{e}\cos{\omega}$             &    & $U(-1, 1)$ & $-0.02 \pm 0.33$ & & $0.07^{+0.21}_{-0.13}$ & \juliet \\
$\sqrt{e}\sin{\omega}$             &    & $U(-1, 1)$ & $-0.02 \pm 0.12$ & & $0.057 \pm 0.090$ & \juliet \\
Dilution g'                        &    & $U(0, 1)$ & $0.84^{+0.11}_{-0.17}$ & & & \\
Dilution TESS sector 22            &    & $U(0, 1)$ & $0.683 \pm 0.071$ & & & \\
Dilution TESS sector 46            &    & $U(0, 1)$ & $0.893 \pm 0.060$ & & & \\
Dilution TESS sector 49            &    & $U(0, 1)$ & $0.880 \pm 0.058$ & &  & \\
Dilution ExTrA                     &    & $U(0, 1)$ & $0.928 \pm 0.057$ & & & \\

\hline
\end{tabular}
\tablefoot{The table lists: prior, posterior median, and 68.3\% CI, for the analysis of transit photometry and RVs (\juliet, in which the asymmetry of the transit is modelled using GPs, and the chromaticity with dilution factors), and the analysis of transit photometry including a starspot in the model (\starry). If a parameter is estimated from both modellings, the last column specifies the adopted value. There is a degeneracy with ($i_\star + 180\degree$) and ($180\degree - \lambda$) angles assuming $i_p$ between 0\degree\ and 180\degree. $^\dagger$ Upper limit, 95\% confidence. The parameters $q_1$ and $q_2$ are the quadratic limb-darkening coefficients parameterised using \citet{kipping2013}. The planetary equilibrium temperature is computed for zero albedo and full day-night heat redistribution. The true spin–orbit angle is given for the prograde and retrograde configurations, which cannot be distinguished with the available data. IAU 2012: \rm{au} = 149$\;$597$\;$870$\;$700~\rm{m}$\;$. IAU 2015: \Rnom = 6.957\ten[8]~\rm{m}, \Renom~=~6.378$\;$1\ten[6]~\rm{m}, \GMnom = 1.327$\;$124$\;$4\ten[20]~$\rm{m^3\;s^{-2}}$, \GMenom = 3.986$\;$004\ten[14]~$\rm{m^3\;s^{-2}}$. \Msun = \GMnom/$\mathcal G$, \MEarth = \GMenom/$\mathcal G$. CODATA 2018: $\mathcal G$ = 6.674$\;$30\ten[-11]~$\rm{m^3\;kg^{-1}\;s^{-2}}$. $N(\mu, \sigma)$: Normal distribution with mean $\mu$, and standard deviation $\sigma$. $U(a, b)$: A uniform distribution defined between a lower $a$ and an upper $b$ limit.}
\end{table*}

\begin{figure}
  \centering
  \includegraphics[width=0.49\textwidth]{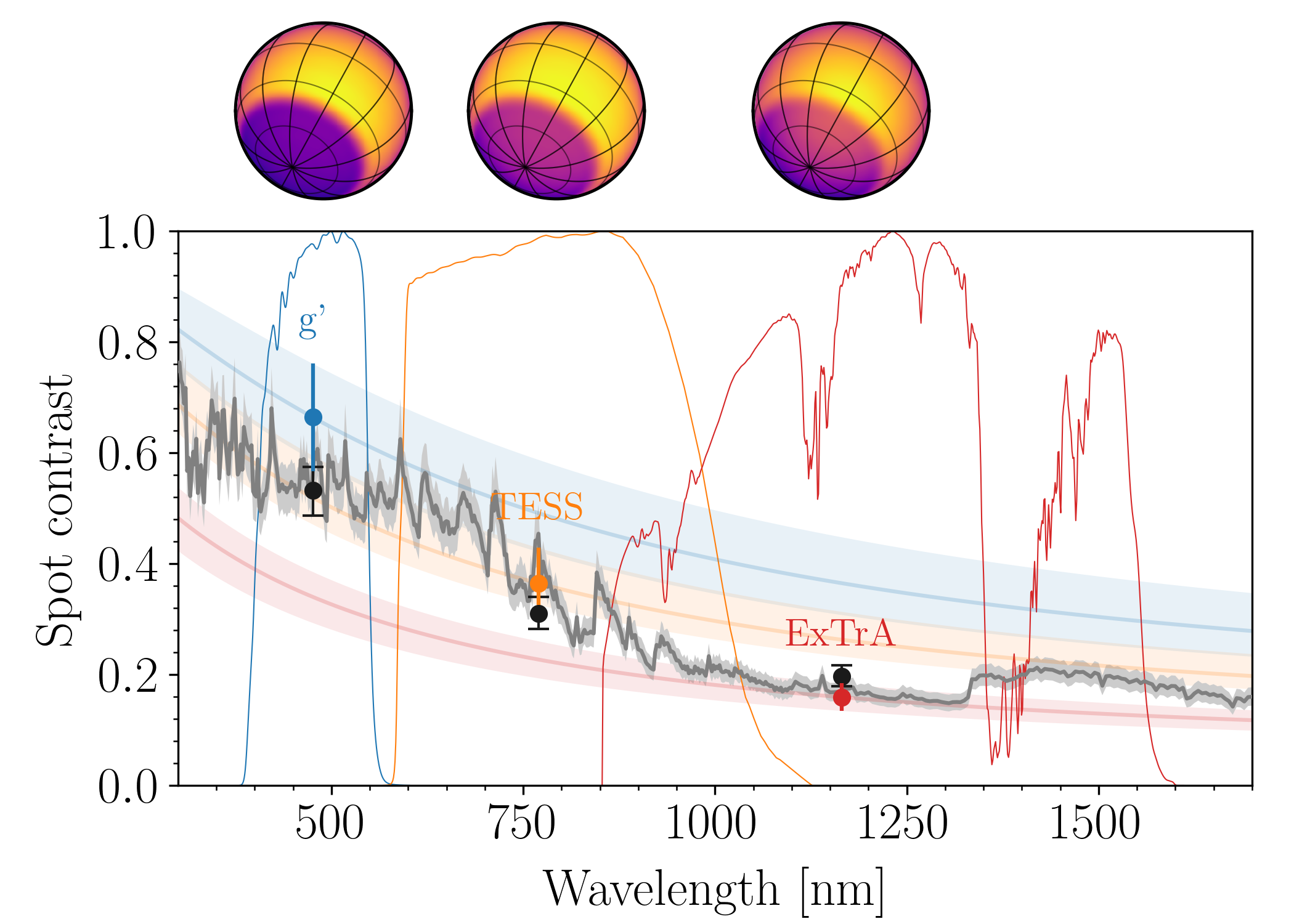}
  \caption{Polar spot contrast as a function of the photometric band (blue for g', orange for TESS, and red for ExTrA) from the analysis of Sect.~\ref{section.analysis}. The g' and TESS transmission curves were retrieved from the SVO Filter Profile Service \citep[\url{http://svo2.cab.inta-csic.es/theory/fps/},][]{rodrigo2012,rodrigo2020}. The blue, orange, and red bands represent the median and the 68\% CI of the contrast as a function of wavelength predicted by the corresponding band contrast measurement assuming blackbody emission. The black error bars and grey curves are the posterior (median and the 68\% CI) of the measured contrast and the contrast as a function of wavelength, respectively, from the modelling using the PHOENIX/BT-Settl stellar atmosphere models (Sect.~\ref{sec.contrast}). {\it Above the panel}: Model of the stellar surface as seen at each band for the median posterior values.} \label{fig.spot_contrast}
\end{figure}

\begin{figure*}
  \centering
  \includegraphics[width=1.0\textwidth]{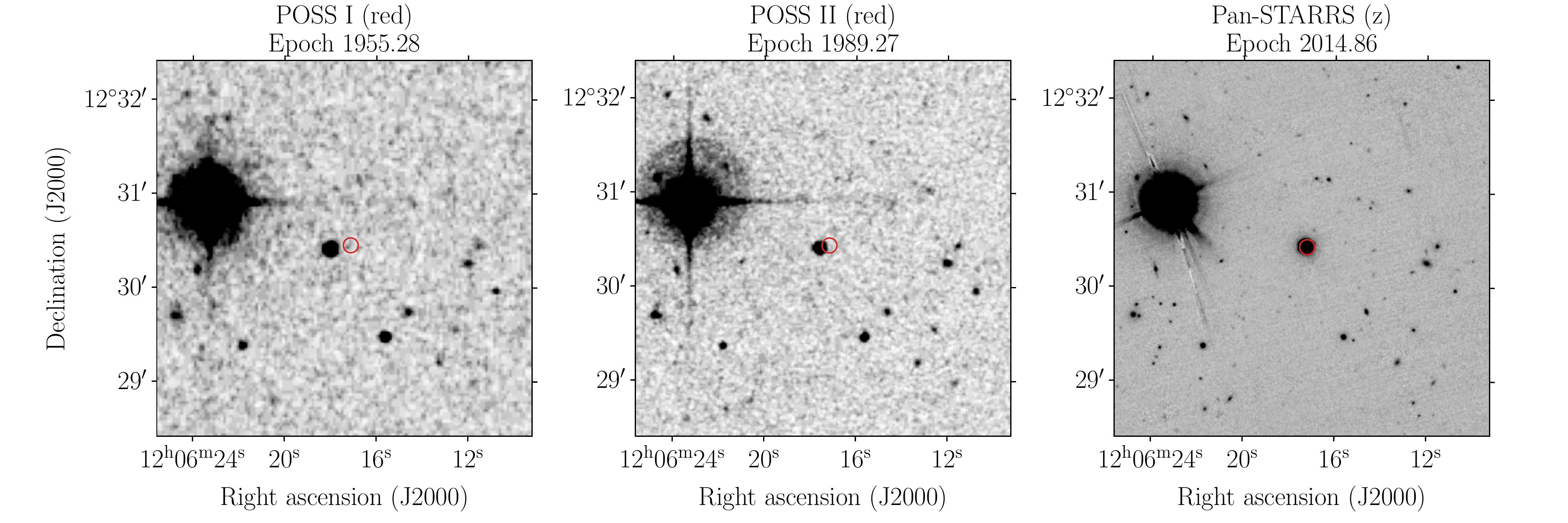}
  \caption{Seeing limited images centred on TOI-3884 at epoch 2016.0 \citep[Gaia EDR3 position,][marked with a red circle]{gaiaEDR3} from the Palomar Observatory Sky Survey (POSS, two epochs, retrieved from \url{https://archive.eso.org/dss/dss}) and Pan-STARRS \citep{chambers2016}. The proper motion can be appreciated over the 60-year time span. No background star is detected at the current position of TOI-3884.} \label{fig.DSS}
\end{figure*}

\begin{figure}
  \centering
  \includegraphics[width=0.49\textwidth]{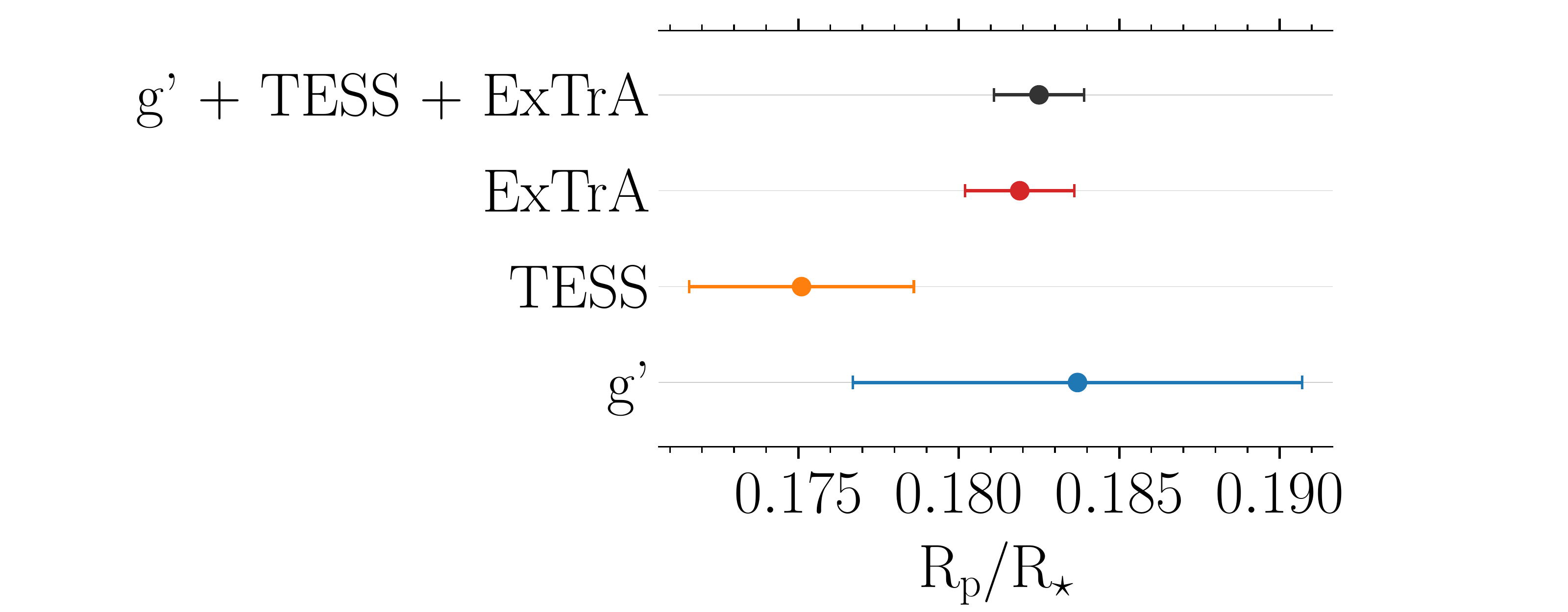}
  \caption{Planet-to-star radius ratio estimated jointly for all bands or for each band individually.} \label{fig.RadiusRatio}
\end{figure}

\FloatBarrier
\section{Spot contrast}\label{sec.contrast}

We modelled the derived spot contrast in three different bands (Table~\ref{table.spot}) to estimate the photosphere-spot difference temperature. The use of a blackbody spectrum fit the data poorly (see Fig.~\ref{fig.spot_contrast}). Therefore, we used the PHOENIX/BT-Settl \citep{allard2012} stellar atmosphere models, which we integrated in the g', TESS, and ExTrA bands. We used as input parameters the effective temperature, surface gravity, and metallicity with normal priors from Sect.~\ref{section.stellar_parameters} (Table~\ref{table.stellar_parameters}), and a photosphere-spot difference temperature with a uniform prior. We sampled from the posterior with \emcee \citep{goodmanweare2010, emcee}. The parameters, priors, and posteriors are listed in Table~\ref{table.spot_contrast}. The posterior model is shown in Fig.~\ref{fig.spot_contrast}. The posterior spot contrast in the $K_s$-band (that was used to derive the stellar parameters in Sect.~\ref{section.stellar_parameters}) is $0.159 \pm 0.017$.

The measurement of the spot contrast in several bands allows for an independent determination of the effective temperature of the star, in addition to the photosphere-spot difference temperature, subject to the validity of the stellar atmosphere model. We repeated the modelling using a uniform prior for the photosphere effective temperature, surface gravity, and metallicity, and obtained $T_{\mathrm{eff}} = 3400^{+460}_{-320}$~K, in agreement with the adopted one (Table~\ref{table.stellar_parameters}), but having a large uncertainty.

\begin{table}[htb]
\tiny
    \renewcommand{\arraystretch}{1.25}
    \setlength{\tabcolsep}{4pt}
\centering
\caption{Modelling of the starspot contrast.}\label{table.spot_contrast}
\begin{tabular}{lccccc}
\hline
\hline
Parameter & & Prior & Posterior median & Prior & Posterior median  \\
&  & & and 68.3\% CI & & and 68.3\% CI\\
\hline
Photosphere effective temperature, $T_{\mathrm{eff}}$ & [K]   & $N$(3269, 70)     & $3249 \pm 74$     & $U$(400, 10000) & $3400^{+460}_{-320}$ \\
Surface gravity, \logg                                & [cgs] & $N$(4.921, 0.028) & $4.922 \pm 0.028$ & $U$(-0.5, 6.0)  & $0.8^{+2.9}_{-1.0}$ \\
Metallicity, $[\rm{Fe/H}]$                            & [dex] & $N$(0.23, 0.12)   & $0.22 \pm 0.12$   & $U$(-0.4, 0.5)  & $-0.87^{+0.92}_{-1.4}$ \\
Photosphere-spot difference temperature               & [K]   & $U$(0, 500)       & $187 \pm 21$      & $U$(0, 500)     & $214^{+95}_{-51}$ \\
\hline
\end{tabular}
\tablefoot{Modelling with informative and non-informative priors for the effective temperature, surface gravity, and metallicity. $N$($\mu$, $\sigma$): Normal distribution prior with mean $\mu$, and standard deviation $\sigma$. $U$(l, u): Uniform distribution prior in the range [l, u].}
\end{table}

\end{appendix}
\end{document}